## Exact value of the correlation factor for the divacancy mechanism in FCC crystals: the end of a long quest

*J.L. Bocquet*
*CMLA, LRC-Méso, ENS CACHAN, 61 Avenue du Président Wilson, 94235 Cachan, France*
*jean-louis.bocquet@cmla.ens-cachan.fr*

## Introduction

Matter transport in crystals is most often mediated through point defects and atomic species are not random walkers. Their walk departs from that of a truly random walk by a factor, generally smaller than unity, called the correlation factor. After a first initial jump has been performed by an atom, a simple inspection of the mechanism at the atomic scale shows that the direction of its next jump has a larger than random probability to happen in the reverse direction. The net result is a decrease in efficiency of the walk.[1] Taking into account the effect of the immediate reverse jump only gives a good order of magnitude of the correlation effect. However the exact value requires a closer examination of the atomic mechanism.

The most simple case ever studied is the diffusion of a tagged atom called 'the tracer' moved by a single defect (vacancy or interstitial) in an infinite crystal lattice. The chemical nature of the tracer can be identical ('self diffusion') or different ('solute diffusion') from that of all other atoms. In both cases the calculation of the correlation effect boils down to evaluate the probabilities $p_{\omega_1,\omega_2}$ of performing a tracer jump $\overrightarrow{\omega_2}$ following a previous jump $\overrightarrow{\omega_1}$, with the help of the *same* defect. The average quadratic displacement of the tracer due to its encounter with a given defect is a function of the number of jumps and of the probabilities $p_{\omega_1,\omega_2}$. Due to the translational symmetry of the medium, the atomic configuration after a tracer jump is identical to that before the jump, apart from a rotation: the probabilities $p_{\omega_1,\omega_2}$ do not depend on $\overrightarrow{\omega_1}$ and $\overrightarrow{\omega_2}$ separately, but only on $\overrightarrow{\omega_2}-\overrightarrow{\omega_1}$. In such an ideal situation, a simple recurrence equation for the displacement of the tracer atom can be written and solved in order to extract the mean quadratic displacement of the tracer due to its encounter with a unique defect. Further, choosing the x-axis as a principal axis of the diffusion tensor restrains the problem to the only consideration of the x-components $x_1$ and $x_2$.[2]

For the monovacancy mechanism, all the jumps have the same x-component 's' and only one type of jump has to be defined. The correlation factor $f_V$ is a simple function of the average product $t_1 = \langle x_1 x_2 \rangle / s^2 = p^+ - p^-$ where $p^+, p^-$ are the probabilities of performing two successive tracer jumps with an x-component of the same sign or not. After the first jump, the defect is in a convenient position to let the tracer atom perform immediately a reverse jump, which explains that $|p^-| > |p^+|$, and $t_1 < 0$. The correlation factor is given by: $f_V = (1+t_1)/(1-t_1)$.

For the more complicated divacancy mechanism, all the jumps have the same x-component but two types of jumps must be defined according to the tracer-divacancy configuration. The correlation factor $f_{2V}$ is a function of the four average products $t_{\alpha\beta} = < x_1^\alpha x_2^\beta >$, where the superscripts $\alpha,\beta$ stand for the type of the jump ($\alpha,\beta = 1,2$). The two types of jump are defined, according to the following rule:

* a type 1 jump is performed if the two vacancies are located on the same x plane before the jump is completed;
* a type 2 jump is performed if only one vacancy is located on the same x plane as that of the tracer before the jump is completed.

In this case, the correlation factor is given by:

$$f_{2V} = \frac{(1+t_{21})(1+t_{12}) - t_{11}t_{22}}{(1-t_{11})(1-t_{22}) - t_{12}t_{21}} \tag{1}$$

Locating the origin of the coordinates on the site occupied by the tracer atom after its first jump $\overrightarrow{\omega_1^\alpha}$ (x-component $x_1^\alpha$) , the problem consists in calculating the total probability for the defect, starting initially at $-\overrightarrow{\omega_1^\alpha}$ to come back in the neighbourhood of the tracer atom at $-\overrightarrow{\omega_2^\beta}$ and make it perform a further jump $\overrightarrow{\omega_2^\beta}$ (x-component $x_2^\beta$).

The precision of the result is a (slowly) varying function of the number of jumps, the defect is allowed to perform before its next exchange with the tracer:

* the first evaluation, made by hand-summing the contributions of short defect trajectories, evaluated the divacancy correlation factor to 0.54; [3]
* the second evaluation amounting to 0.475 rested on a matrix method and took account of divacancy trajectories of arbitrary length, provided they were confined in a small volume V around the tracer atom;[4]
* the third evaluation, using an extended version of the matrix method (i.e. a larger V) decreased the value down to 0.472;[5]
* the best values known up to now were later provided by direct Monte-Carlo simulations: 0.458 ± 0.001 [6] and 0.4582 ± 0.0005. [7]

The object of the present contribution is to determine the first exact evaluation of this quantity with an integral method already illustrated in previous publications.[8-10] The first part is devoted to notations and definitions. The second part establishes the transport equations for the divacancy mechanism. The third part solves the linear system yielding the average products $t_{\alpha\beta}$ together with the final exact value of the correlation factor $f_{2V}$. Details of calculations are passed in Appendices:

*Appendix A gives useful expressions of the determinant and cofactors entering the solution;
* Appendix B determines the coefficients of the linear system yielding the return probabilities of the vacancy on sites giving rise to a next tracer jump;
* Appendices C and D establish the formal expressions of coefficients entering the definition of $t_{\alpha\beta}$;
* Appendix E establishes the equality $t_{11} = t_{22}$.

## 1. Definitions and notations

The sites in the host FCC lattice are denoted by: $\Omega_{ix+jy+kz} = i\,\Omega_x + j\,\Omega_y + k\,\Omega_z$ where i, j, k are integers complying with the condition i+j+k even and the basis vectors are defined according to $\Omega_x = \frac{a}{2}[1,0,0]$ $\Omega_y = \frac{a}{2}[0,1,0]$ $\Omega_z = \frac{a}{2}[0,0,1]$, with « a » standing for the FCC lattice parameter.

The centre of the divacancy defines unambiguously the position of the two vacancies: it moves on a new lattice which is made up by the middles of all nearest neighbour bonds of the parent FCC lattice. The new sites will be described with the help of new basis vectors defining the new unit length along the three coordinate axis: $\omega_x = \frac{a}{4}[1,0,0]$, $\omega_y = \frac{a}{4}[0,1,0]$, $\omega_z = \frac{a}{4}[0,0,1]$. The orientation axis of the divacancy is given by the direction of the line segment joining the two vacancies of the pair.

Three different types of sites are to be considered and denoted with a subscript 'x' , 'y' or 'z', the latter being related to the plane containing the two ends of the divacancy:

* 'x' sites correspond to divacancies contained in planes $x = 2i$; the 'x' sites are located at $r_{2ix+(2j\pm1)y+(2k\pm1)z} = 2i\ \omega_x + (2j \pm 1)\ \omega_y + (2k \pm 1)\ \omega_z$. The orientation of the divacancy is a function of the parity of 'i'. If 'i' is even, the orientation of the divacancy is oriented along $\omega_{y+z}$ for sites $r_{2ix+(2j+1)y+(2k+1)z}$ and $r_{2ix+(2j-1)y+(2k-1)z}$, and oriented along $\omega_{y-z}$ for sites $r_{2ix+(2j+1)y+(2k-1)z}$ and $r_{2ix+(2j-1)y+(2k+1)z}$. The reverse is true when 'i' is odd. The first neighbours of 'x' sites belonging to plane $x = 2i$ lie only in adjacent $x = 2i+1$ and $x = 2i-1$ planes and are exclusively of 'y' or 'z' type;

* 'y' sites correspond to divacancies contained in plane $y = 2j$; the 'y' sites are located at $r_{(2i\pm1)x+2jy+(2k\pm1)z} = (2i \pm 1)\ \omega_x + 2j\ \omega_y + (2k \pm 1)\ \omega_z$; the orientation of the divacancy follows similar rules as above after a cyclic permutation $i \to j,\ j \to k,\ k \to i$ but now depends on the parity of 'j'. If 'j' is even, the orientation of the divacancy is oriented along $\omega_{z+x}$ for sites $r_{(2i+1)x+2jy+(2k+1)z}$ and $r_{(2i-1)x+2jy+(2k-1)z}$, and oriented along $\omega_{z-x}$ for sites $r_{(2i+1)x+2jy+(2k-1)z}$ and $r_{(2i-1)x+2jy+(2k+1)z}$. The reverse is true when 'j' is odd. The first neighbours of 'y' sites belonging to plane $y = 2j$ lie only in adjacent $y = 2j+1$ and $y = 2j-1$ planes and are exclusively of 'z' or 'x' type;

* 'z' sites correspond to divacancies contained in plane $z = 2k$; the 'z' sites are located at $r_{(2i\pm1)x+(2j\pm1)y+2kz} = (2i \pm 1)\ \omega_x + (2j \pm 1)\ \omega_y + 2k\omega_z$; the orientation of the divacancy depends on the parity of 'k'. If 'k' is even, the orientation of the divacancy is oriented along $\omega_{x+y}$ for sites $r_{(2i+1)x+(2j+1)y+2kz}$ and $r_{(2i-1)x+(2j-1)y+2kz}$, and oriented along $\omega_{x-y}$ for sites $r_{(2i+1)x+(2j-1)y+2kz}$ and $r_{(2i-1)x+(2j+1)y+2kz}$. The reverse is true when 'k' is odd. The first neighbours of 'z' sites belonging to plane $z = 2k$ lie only in adjacent $z = 2k+1$ and $z = 2k-1$ planes and are exclusively of 'x' or 'y' type.

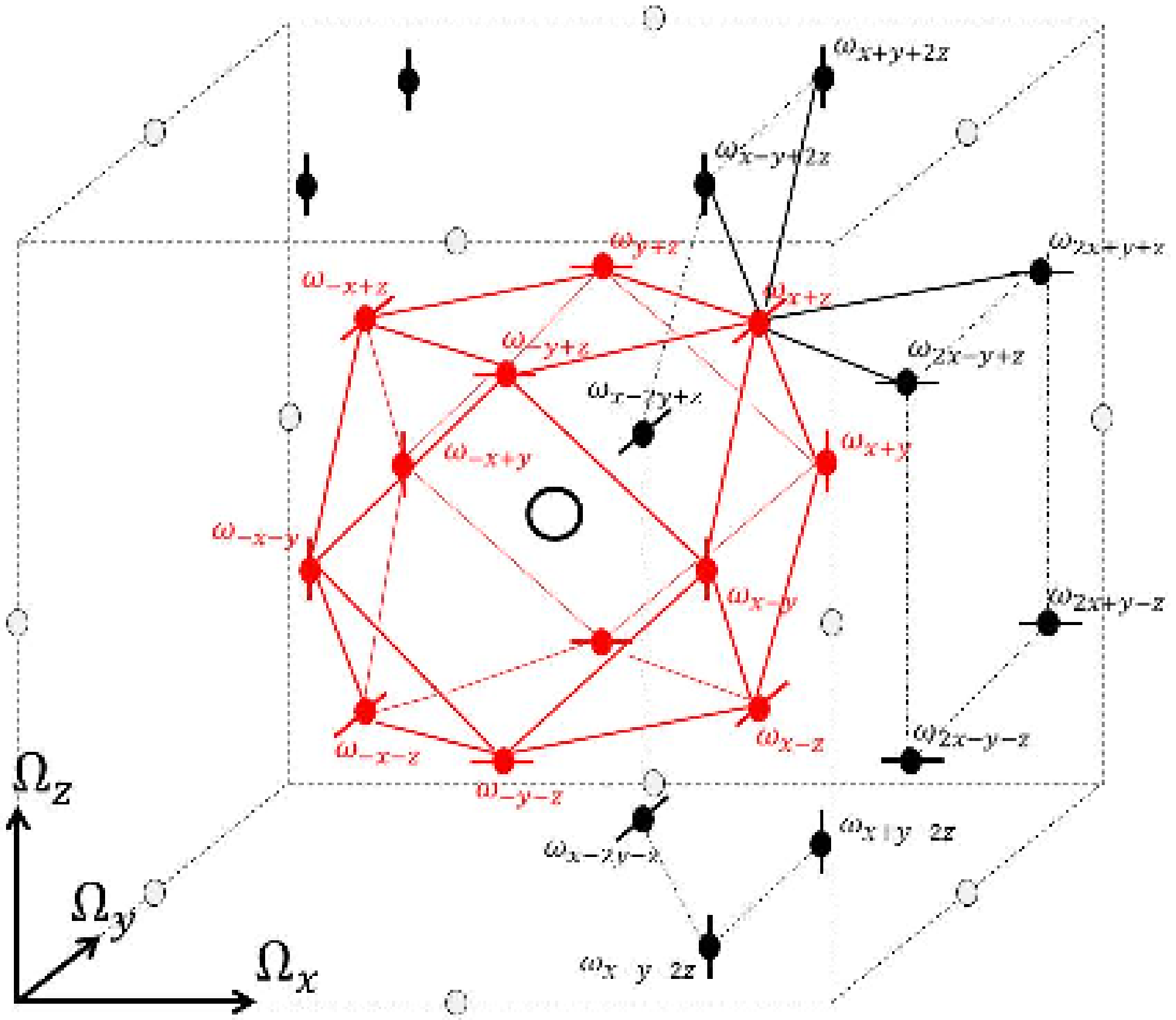

*Fig. 1. The empty circles denote the sites of the host FCC lattice. The filled circles denote the new lattice sites swept by the divacancy centre. Sites coloured in red represent the sink sites around the origin. The continuous thin lines (red or back) represent the possible first neighbour transitions between new sites. The eight first neighbours of 'y' site* $\omega_{x+z}$ *are displayed.*

The current vector in the new lattice will mostly be denoted by the symbol 'r'; '$\omega$' will be used for a small number of sites around the origin, which play a particular role in the problem.

A picture showing the position of the new sites with respect to FCC lattice sites is displayed in Fig. 1. Each site of type 'x', 'y' or 'z' is represented by a filled circle attached to a small line segment aligned along the x, y, or z axis respectively. An alternative view of this new lattice, without any reference to the host FCC lattice, is displayed in Fig. 2: it consists of a simple cubic array of octahedrons sharing their vertices with a coordination number equal to 8.

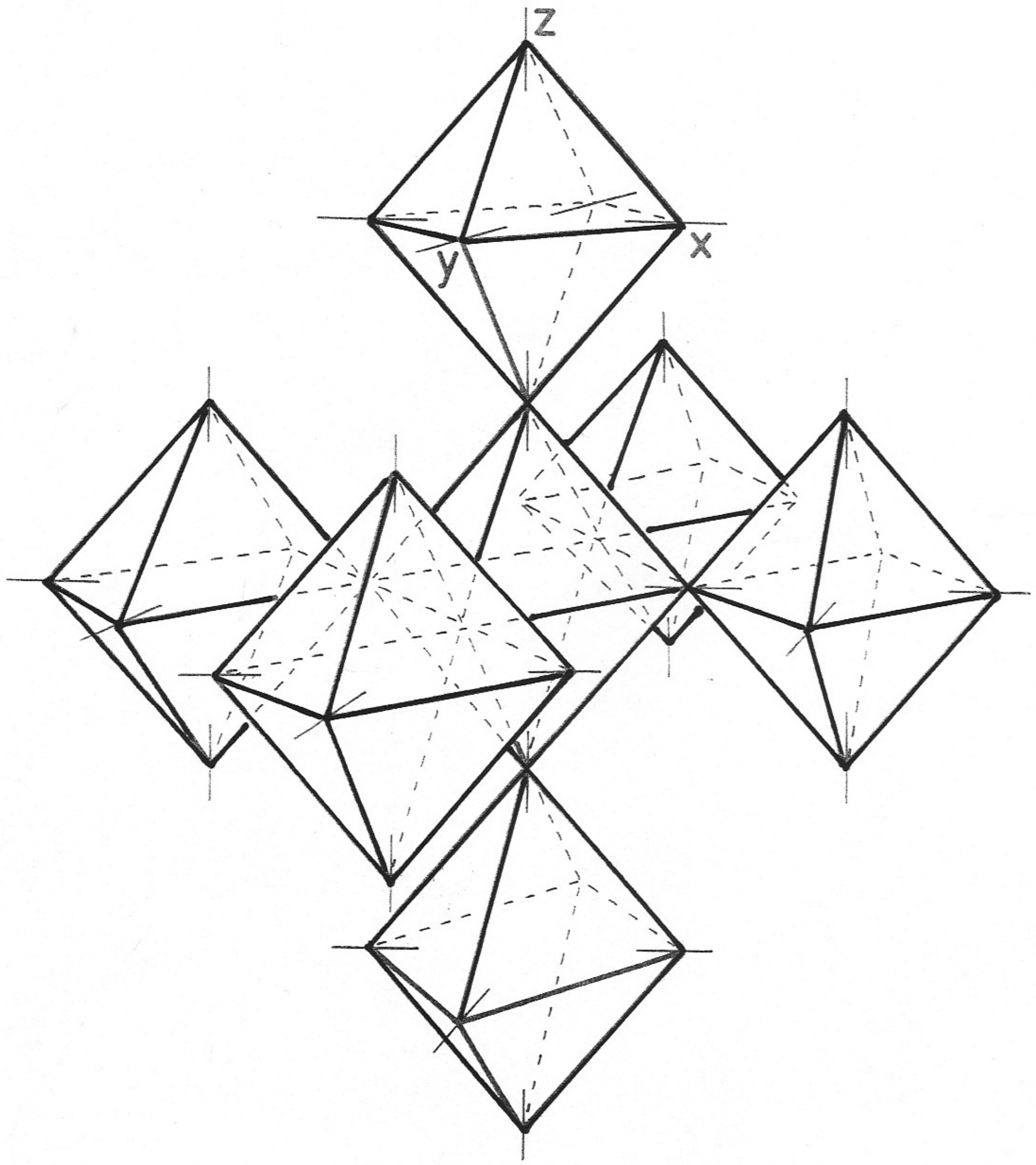

*Fig. 2. New lattice of coordinance 8 swept by the divacancy centre.*

Since the divacancy is not allowed to dissociate, the two constituting vacancies can only exchange with one of the four first neighbours they have in common on the FCC host lattice; hence a total of eight possible jumps for the migration of the complex.

Three density probability functions $L_x^{(\upsilon)}(r,t),\ L_y^{(\upsilon)}(r,t),\ L_z^{(\upsilon)}(r,t)$ will be introduced for the probability of finding a vacancy on the three types of site at time t; the superscript $(\upsilon)$ keeps track of the initial condition, i.e. the location of the divacancy after the first tracer jump of type $(\upsilon)$. A picture of the two types of jump is displayed on Fig. 3.

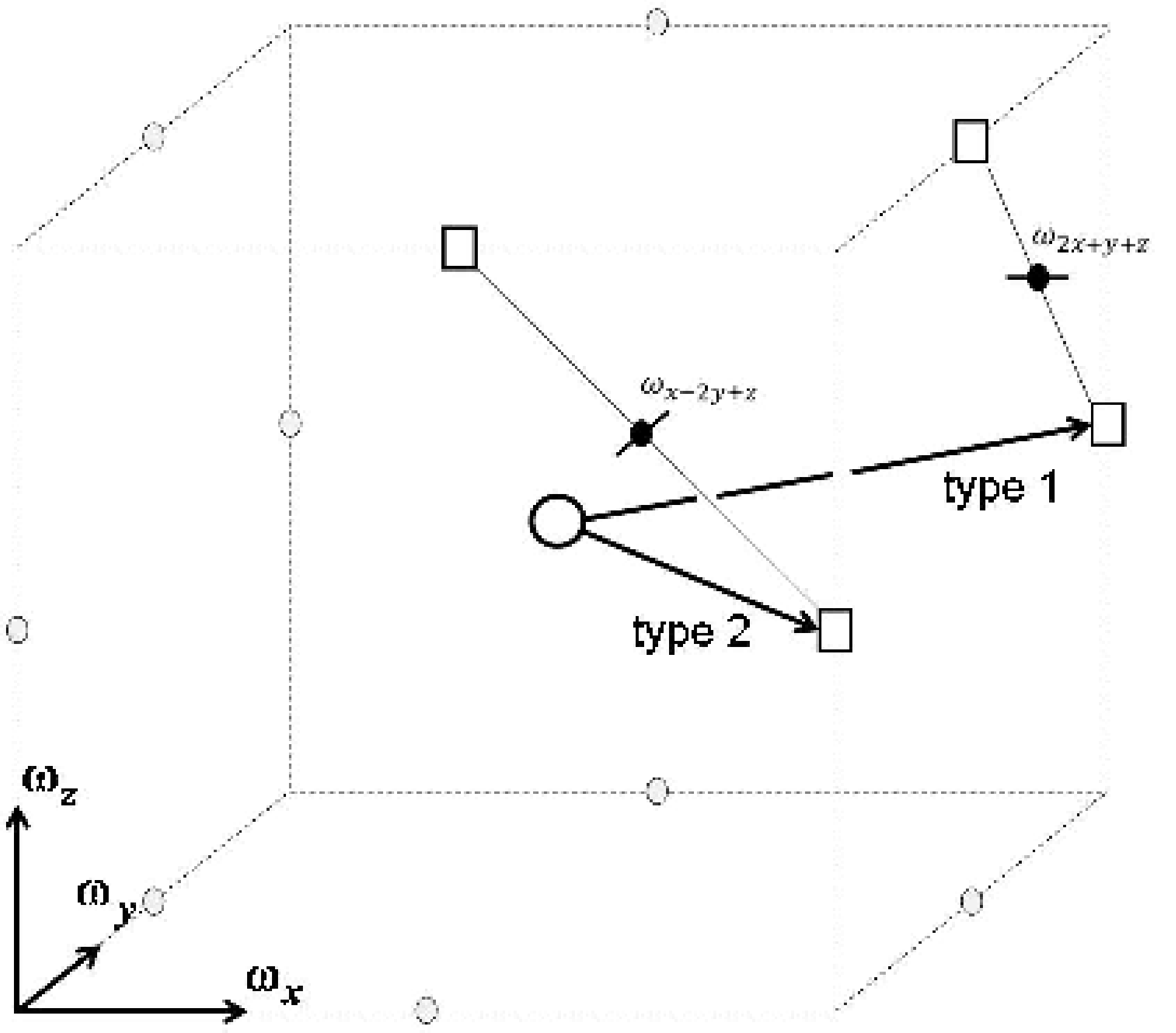

*Fig. 3. The empty circles denote the sites of the host FCC lattice. Squares denote the position of the vacancies. The filled circles denote the sites of the corresponding divacancy centre. The arrows with thick lines denote jumps available for the tracer located on the origin.*

The statement of the correlation problem then runs as follows: assuming that the tracer atom just performed a jump of type $(\upsilon)$ with a negative x-component and taking the new tracer site as the origin of the coordinates, the divacancy is modelled by a unit source which is put on its starting site and spreads in time according to diffusion laws. The purpose of the calculation is to determine which fraction of the unit source flows onto each of the sites close to the origin, namely the four 'x' sites $\omega_{y+z}, \omega_{y-z}, \omega_{-y+z}, \omega_{-y-z}$ of plane x=0, the four 'y' sites $\omega_{z+x}, \omega_{z-x}, \omega_{-z+x}, \omega_{-z-x}$ of plane $y=0$ and the four 'z' sites $\omega_{x+y}, \omega_{x-y}, \omega_{-x+y}, \omega_{-x-y}$ of plane $z=0$. The arrival of the vacancy centre on any one of these sites gives rise to a further tracer jump; they play the role of sinks in the calculation of the return probabilities and are characterized by a total escape frequency equal to zero.

For instance, the fraction of the unit source flowing onto the sink site $\omega_{x+z}$ is given by the time integral of the density probability $L_c^{(\upsilon)}(\omega_{x+z}+\omega_{1V},t)$ times the divacancy jump rate $W_O$, where $\omega_{x+z}+\omega_{1V}$ is a first neighbour of type 'c' of $\omega_{x+z}$ i.e.

$$W_O\int_0^\infty L_c^{(\upsilon)}(\omega_{x+z}+\omega_{1V},t)dt = W_O LL_c^{(\upsilon)}(\omega_{x+z}+\omega_{1V},p)\Big|_{p=0}, \quad (2)$$

where $LL_c^{(\upsilon)}$ stands for the Laplace transforms of $L_c^{(\upsilon)}$.

If the divacancy centre is located on sites $\omega_{x+z}+\omega_{1V}=\omega_{2x+y+z},\omega_{2x-y+z}$ the type of the next tracer jump is equal to 1; if it is located on sites $\omega_{x+y+2z},\omega_{x-y+2z}$, the type of the next tracer jump is equal to 2.
Summing Eq. 2 on all sink sites gives the total fraction of the unit source flowing back onto the tracer atom. In 3D lattices, this fraction is strictly smaller than unity, since the defect can escape to infinity without any further exchange jump than the first.
The isotropy of the matter transport by this mechanism together with the symmetries of the new lattice allow us to use initial conditions (CI) with a four-fold symmetry around the x-axis and a mirror symmetry plane at x=0, as already illustrated in more detail in a previous work devoted to the monovacancy mechanism.[10]

### 1.1 Initial condition CI1

A previous type 1 jump has been performed with a negative x-component: this means that, before the jump, the divacancy was in plane x=0 at any of the sites $\omega_{y+z},\omega_{y-z},\omega_{-y+z},\omega_{-y-z}$ and the tracer was located at one of the four possible sites $\omega_{2x+2y},\omega_{2x-2z},\omega_{2x+2z},\omega_{2x-2y}$. After completion of this first jump, the number of resulting possible configurations amounts to eight, with a divacancy located at four 'y' sites $\omega_{x+2y+z},\omega_{x+2y-z},\omega_{x-2y+z},\omega_{x-2y-z}$ or four 'z' sites $\omega_{x+y+2z},\omega_{x+y-2z},\omega_{x-y+2z},\omega_{x-y-2z}$. The unit source of the diffusion problem is thus replaced by eight sources with a weight +1/8 on the four 'y' sites and the four 'z' sites of plane $x=+1$. The mirror source -1 is replaced by eight sources with weight -1/8 on the mirror-symmetric sites on plane $x=-1$, that is the four 'y' sites $\omega_{-x-2y-z},\omega_{-x-2y+z},\omega_{-x+2y-z},\omega_{-x+2y+z}$ and the four 'z' sites $\omega_{-x-y-2z},\omega_{-x-y+2z},\omega_{-x+y-2z},\omega_{-x+y+2z}$.

In all what follows a shorthand notation will replace $\cos(lk_x+mk_y+nk_z)$ and $\sin(lk_x+mk_y+nk_z)$ by $c_{lx+my+nz}$ and $s_{lx+my+nz}$ respectively. The indices $l,m,n$ will not be mentionned when equal to unity. The initial condition for 'y' sites and its Fourier transform are then given by:

$$\begin{aligned} L_y^{(1)}(r,t=0)=&\frac{1}{8}\left[\delta(r-\omega_{x+2y+z})+\delta(r-\omega_{x+2y-z})+\delta(r-\omega_{x-2y+z})+\delta(r-\omega_{x-2y-z})\right]\\ &-\frac{1}{8}\left[\delta(r-\omega_{-x+2y+z})+\delta(r-\omega_{-x+2y-z})+\delta(r-\omega_{-x-2y+z})+\delta(r-\omega_{-x-2y-z})\right] \qquad (3.1)\\ \rightarrow FL_y^{(1)}=&-ic_{2y}c_zs_x \end{aligned}$$

where the superscript refers to the type of the previous jump. The Fourier transform of $L_y^{(1)}(r,t)$ is defined by $FL_y^{(1)}(k,t)=\sum_r e^{-ikr}L_y^{(1)}(r,t)$ where the summation runs over all sites swept by the centre of the divacancy and $kr$ stands for a scalar product.

Due to the four-fold symmetry, the z variable plays the same role as y, hence:

$$L_z^{(1)}(r,t=0)=\frac{1}{8}\left[\delta(r-\omega_{x+2y+z})+\delta(r-\omega_{x+2y-z})+\delta(r-\omega_{x-2y+z})+\delta(r-\omega_{x-2y-z})\right]$$
$$-\frac{1}{8}\left[\delta(r-\omega_{-x+2y+z})+\delta(r-\omega_{-x+2y-z})+\delta(r-\omega_{-x-2y+z})+\delta(r-\omega_{-x-2y-z})\right] \quad (3.2)$$
$$\rightarrow FL_z^{(1)}=-ic_yc_{2z}s_x$$

Together with the initial condition for $L_x^{(1)}(r,t)$:

$$\rightarrow FL_x^{(1)}=0. \quad (3.3)$$

### 1.2 Initial condition CI2

A previous type 2 jump has been performed with a negative x-component: this means that before the jump the divacancy was located at $\omega_{x+2y+z}$ $\omega_{x+2y-z}$ $\omega_{x-2y+z}$ $\omega_{x-2y-z}$ or $\omega_{x+y+2z}$, $\omega_{x+y-2z}$, $\omega_{x-y+2z}$, $\omega_{x-y-2z}$ and the tracer was located at one of the four possible sites $\omega_{2x+2y}, \omega_{2x-2z}, \omega_{2x+2z}, \omega_{2x-2y}$. The resulting configurations amount to four with a tracer atom on the origin and the divacancy located on sites of type 'x' at $\omega_{2x+y+z}$, $\omega_{2x-y+z}$, $\omega_{2x+y-z}$, $\omega_{2x-y-z}$. The unit source +1 of the diffusion problem is thus split into four sources with weight +1/4 on the four 'x' sites of plane $x=+2$ located at $\omega_{2x+y+z}$, $\omega_{2x-y+z}$, $\omega_{2x+y-z}$, $\omega_{2x-y-z}$; the mirror source -1 is split into four sources with weight -1/4 on the four 'x' sites of plane $x=-2$ at $\omega_{-2x+y+z}$, $\omega_{-2x-y+z}$, $\omega_{-2x+y-z}$, $\omega_{-2x-y-z}$.

The Fourier transform of the initial condition for the function $L_x^{(2)}(r,t)$ is given by:

$$FL_x^{(2)}=\frac{1}{4}\left[\begin{array}{l} e^{-ik\omega_{2x+y+z}}+e^{-ik\omega_{2x+y-z}}+e^{-ik\omega_{2x-y+z}}+e^{-ik\omega_{2x-y-z}} \\ \qquad -e^{-ik\omega_{-2x+y+z}}-e^{-ik\omega_{-2x+y-z}}-e^{-ik\omega_{-2x-y+z}}-e^{-ik\omega_{-2x-y-z}}\end{array}\right] \quad (4.1)$$
$$=-4ic_xc_yc_zs_x$$

together with the Fourier transformed initial conditions for functions $L_y^{(2)}(r,t)$, $L_z^{(2)}(r,t)$:

$$\rightarrow FL_y^{(2)}=0 \quad (4.2)$$

$$\rightarrow FL_z^{(2)}=0 \quad (4.3)$$

### 1.3 Relationships between unknowns due to symmetries

The mirror symmetry plane leads to presence probabilities equal to zero on all sites of plane $x=0$, and to presence probabilities which are equal in magnitude but of opposite signs on the sites of planes $x=+i$ and $x=-i$.

The presence probabilities on sink sites, which are first neighbours of the origin, are named with particular subscripts $_{x0}$, $_{y1}$, $_{z1}$:

$$L_x^{(\upsilon)}(\omega_{y+z},t) = L_x^{(\upsilon)}(\omega_{y-z},t) = L_x^{(\upsilon)}(\omega_{-y+z},t) = L_x^{(\upsilon)}(\omega_{-y-z},t) = L_{xo}^{(\upsilon)} = 0$$
$$L_y^{(\upsilon)}(\omega_{x+z},t) = L_y^{(\upsilon)}(\omega_{x-z},t) = L_{y1}^{(\upsilon)} \qquad L_y^{(\upsilon)}(\omega_{-x+z},t) = L_y^{(\upsilon)}(\omega_{-x-z},t) = -L_{y1}^{(\upsilon)} \tag{5}$$
$$L_z^{(\upsilon)}(\omega_{x+y},t) = L_z^{(\upsilon)}(\omega_{x-y},t) = L_{y1}^{(\upsilon)} \qquad L_z^{(\upsilon)}(\omega_{-x+y},t) = L_z^{(\upsilon)}(\omega_{-x-y},t) = -L_{z1}^{(\upsilon)}$$
with $L_{y1}^{(\upsilon)} = L_{z1}^{(\upsilon)}$.

In the same way, the 'y' and 'z' sites in the planes $x = +1$ and $x = -1$ which are first neighbours of sinks without being sinks themselves play a particular role in the diffusion problem. They are named with particular subscripts $_{y2}$, $_{z2}$:

$$L_y^{(\upsilon)}(\omega_{x+2y+z},t) = L_y^{(\upsilon)}(\omega_{x-2y+z},t) = L_y^{(\upsilon)}(\omega_{x+2y-z},t) = L_y^{(\upsilon)}(\omega_{x-2y-z},t) = L_{y2}^{(\upsilon)}$$
$$L_y^{(\upsilon)}(\omega_{-x+2y+z},t) = L_y^{(\upsilon)}(\omega_{-x-2y+z},t) = L_y^{(\upsilon)}(\omega_{-x+2y-z},t) = L_y^{(\upsilon)}(\omega_{-x-2y-z},t) = -L_{y2}^{(\upsilon)} \tag{6}$$
$$L_z^{(\upsilon)}(\omega_{x+y+2z},t) = L_z^{(\upsilon)}(\omega_{x-y+2z},t) = L_z^{(\upsilon)}(\omega_{x+y-2z},t) = L_z^{(\upsilon)}(\omega_{x-y-2z},t) = L_{z2}^{(\upsilon)}$$
$$L_z^{(\upsilon)}(\omega_{-x+y+2z},t) = L_z^{(\upsilon)}(\omega_{-x-y+2z},t) = L_z^{(\upsilon)}(\omega_{-x+y-2z},t) = L_z^{(\upsilon)}(\omega_{-x-y-2z},t) = -L_{z2}^{(\upsilon)}$$
with $L_{y2}^{(\upsilon)} = L_{z2}^{(\upsilon)}$.

The time integral of the probabilities on sink sites are given by their Laplace transforms calculated with $p = 0$; they are denoted S with the same subscripts and superscripts as above and verify the following relationships:

$$S_{xo}^{(\upsilon)} = \int_{t=0}^{\infty} L_x^{(\upsilon)}(\omega_{y+z},t)dt = LL_{xo}^{(\upsilon)}(p)\Big|_{p=0} = \frac{1}{V_{ZB}}\int FLL_x^{(\upsilon)}(k,p)\Big|_{p=0} e^{+ik\omega_{y+z}} d_3k = 0$$
$$S_{y1}^{(\upsilon)} = \int_{t=0}^{\infty} L_y^{(\upsilon)}(\omega_{x+z},t)dt = LL_{y1}^{(\upsilon)}(p)\Big|_{p=0} = \frac{1}{V_{ZB}}\int FLL_y^{(\upsilon)}(k,p)\Big|_{p=0} e^{+ik\omega_{x+z}} d_3k \tag{7}$$
$$S_{z1}^{(\upsilon)} = \int_{t=0}^{\infty} L_z^{(\upsilon)}(\omega_{x+y},t)dt = LL_{z1}^{(\upsilon)}(p)\Big|_{p=0} = \frac{1}{V_{ZB}}\int FLL_z^{(\upsilon)}(k,p)\Big|_{p=0} e^{+ik\omega_{x+y}} d_3k$$

with $S_{y1}^{(\upsilon)} = S_{z1}^{(\upsilon)}$ because of the four-fold symmetry, $V_{ZB}$ the first Brillouin zone and $FLL_x^{(\upsilon)}(k,p)$, $FLL_y^{(\upsilon)}(k,p)$, $FLL_z^{(\upsilon)}(k,p)$ the Fourier-Laplace transforms of the functions $L_x^{(\upsilon)}(r,t)$, $L_y^{(\upsilon)}(r,t)$, $L_z^{(\upsilon)}(r,t)$ with initial conditions $(\upsilon)$. Thanks to mirror symmetry:

$$\int_{t=0}^{\infty} L_y^{(\upsilon)}(\omega_{-x+z},t)dt = -S_{y1}^{(\upsilon)} \qquad \int_{t=0}^{\infty} L_z^{(\upsilon)}(\omega_{-x+y},t)dt = -S_{z1}^{(\upsilon)}$$

The quantity $S_{y1}^{(\upsilon)} = S_{z1}^{(\upsilon)}$ increases in a monotonic way towards an horizontal asymptote at long times: its Laplace transform will diverge as $1/p$ when $p \to 0$. We will see in Appendix B that $S_{y1}^{(\upsilon)}$ always appears in the equations as a product $pS_{y1}^{(\upsilon)}$ which tends towards a finite value when $p \to 0$.

For the sites in the planes $x = +1$ and $x = -1$ which are first neighbours of sinks without being sinks themselves:

$$LL_{y2}^{(\upsilon)} = \int_{t=0}^{\infty} L_y^{(\upsilon)}(\omega_{x+2y+z}, t)dt = LL_{y2}^{(\upsilon)}(p)\Big|_{p=0} = \frac{1}{V_{ZB}} \int FLL_y^{(\upsilon)}(k,p)\Big|_{p=0} e^{+ik\omega_{x+2y+z}} d_3k$$
$$LL_{z2}^{(\upsilon)} = \int_{t=0}^{\infty} L_z^{(\upsilon)}(\omega_{x+y+2z}, t)dt = LL_{z2}^{(\upsilon)}(p)\Big|_{p=0} = \frac{1}{V_{ZB}} \int FLL_z^{(\upsilon)}(k,p)\Big|_{p=0} e^{+ik\omega_{x+y+2z}} d_3k \quad (8)$$

with $S_{y2}^{(\upsilon)} = S_{z2}^{(\upsilon)}$. Thanks to mirror symmetry:

$$\int_{t=0}^{\infty} L_y^{(\upsilon)}(\omega_{-x+2y+z}, t)dt = -LL_{y2}^{(\upsilon)} \qquad \int_{t=0}^{\infty} L_z^{(\upsilon)}(\omega_{-x+y+2z}, t)dt = -LL_{z2}^{(\upsilon)}$$

Unlike the quantity $S_{y1}^{(\upsilon)} = S_{z1}^{(\upsilon)}$, the time integrals $LL_{y2}^{(\upsilon)} = LL_{z2}^{(\upsilon)}$ remain finite when $p \to 0$.

## 2. Transport equations

### 2.1 Transport equation for the function $L_x^{(\upsilon)}(r,t)$

The following three lines correspond to the general term $GT_x^{(\upsilon)}$ describing the exchanges among the probability functions which are stem from first neighbour jumps:

$$\frac{dL_x^{(\upsilon)}(r,t)}{dt} = \left.\begin{matrix} -8W_O L_x^{(\upsilon)}(r,t) \\ +W_O\left[L_y^{(\upsilon)}(r+\omega_{x+y},t) + L_y^{(\upsilon)}(r+\omega_{x-y},t) + L_y^{(\upsilon)}(r+\omega_{-x+y},t) + L_y^{(\upsilon)}(r+\omega_{-x-y},t)\right] \\ +W_O\left[L_z^{(\upsilon)}(r+\omega_{x+z},t) + L_z^{(\upsilon)}(r+\omega_{x-z},t) + L_z^{(\upsilon)}(r+\omega_{-x+z},t) + L_z^{(\upsilon)}(r+\omega_{-x-z},t)\right] \end{matrix}\right\} GT_x^{(\upsilon)}$$

The right hand side must then be corrected to take into account the specificity of sinks sites: the total escape frequency from the sink sites is zero and the inward contribution of these sinks to neighbouring sinks must be subtracted:

$$+\delta(r-\omega_{y+z})\begin{pmatrix} 8W_O L_x^{(\upsilon)}(\omega_{y+z},t) - W_O L_y^{(\upsilon)}(\omega_{x+z},t) - W_O L_z^{(\upsilon)}(\omega_{x+y},t) \\ -W_O L_y^{(\upsilon)}(\omega_{-x+z},t) - W_O L_z^{(\upsilon)}(\omega_{-x+y},t) \end{pmatrix}$$
$$+\delta(r-\omega_{y-z})\begin{pmatrix} 8W_O L_x^{(\upsilon)}(\omega_{y-z},t) - W_O L_y^{(\upsilon)}(\omega_{x-z},t) - W_O L_z^{(\upsilon)}(\omega_{x+y},t) \\ -W_O L_y^{(\upsilon)}(\omega_{-x-z},t) - W_O L_z^{(\upsilon)}(\omega_{-x+y},t) \end{pmatrix}$$
$$+\delta(r-\omega_{-y+z})\begin{pmatrix} 8W_O L_x^{(\upsilon)}(\omega_{-y+z},t) - W_O L_y^{(\upsilon)}(\omega_{x+z},t) - W_O L_z^{(\upsilon)}(\omega_{x-y},t) \\ -W_O L_y^{(\upsilon)}(\omega_{-x+z},t) - W_O L_z^{(\upsilon)}(\omega_{-x-y},t) \end{pmatrix}$$
$$+\delta(r-\omega_{-y-z})\begin{pmatrix} 8W_O L_x^{(\upsilon)}(\omega_{-y-z},t) - W_O L_y^{(\upsilon)}(\omega_{x-z},t) - W_O L_z^{(\upsilon)}(\omega_{x-y},t) \\ -W_O L_y^{(\upsilon)}(\omega_{-x-z},t) - W_O L_z^{(\upsilon)}(\omega_{-x-y},t) \end{pmatrix}$$

Then a correction is needed for the four 'x' sites in planes $x=+2$ and $x=-2$, which are first neighbours of sink sites, without being sinks themselves. For all of them the contribution coming from sink sites must be deleted:

$$
\begin{aligned}
&+\delta(r-\omega_{2x+y+z})\left[-W_O L_y^{(\upsilon)}(\omega_{x+z},t)-W_O L_z^{(\upsilon)}(\omega_{x+y},t)\right]\\
&+\delta(r-\omega_{-2x+y+z})\left[-W_O L_y^{(\upsilon)}(\omega_{-x+z},t)-W_O L_z^{(\upsilon)}(\omega_{-x+y},t)\right]\\
&+\delta(r-\omega_{2x-y+z})\left[-W_O L_y^{(\upsilon)}(\omega_{x+z},t)-W_O L_z^{(\upsilon)}(\omega_{x-y},t)\right]\\
&+\delta(r-\omega_{-2x-y+z})\left[-W_O L_y^{(\upsilon)}(\omega_{-x+z},t)-W_O L_z^{(\upsilon)}(\omega_{-x-y},t)\right]\\
&+\delta(r-\omega_{2x+y-z})\left[-W_O L_y^{(\upsilon)}(\omega_{x-z},t)-W_O L_z^{(\upsilon)}(\omega_{x+y},t)\right]\\
&+\delta(r-\omega_{-2x+y-z})\left[-W_O L_y^{(\upsilon)}(\omega_{-x-z},t)-W_O L_z^{(\upsilon)}(\omega_{-x+y},t)\right]\\
&+\delta(r-\omega_{2x-y-z})\left[-W_O L_y^{(\upsilon)}(\omega_{x-z},t)-W_O L_z^{(\upsilon)}(\omega_{x-y},t)\right]\\
&+\delta(r-\omega_{-2x-y-z})\left[-W_O L_y^{(\upsilon)}(\omega_{-x-z},t)-W_O L_z^{(\upsilon)}(\omega_{-x-y},t)\right]
\end{aligned}
$$

The complete equation is Fourier and Laplace transformed according to

$$
FL_x^{(\upsilon)}(k,t)=\sum_{\{r\}} e^{-ikr}L_x^{(\upsilon)}(r,t) \quad \rightarrow \sum_{\{r\}} e^{-ikr}L_x^{(\upsilon)}(r+\omega,t)=FL_x^{(\upsilon)}(k,t)e^{ik\omega}
$$

$$
\int_0^\infty FL_x^{(\upsilon)}(k,t)e^{-pt}dt = FLL_x^{(\upsilon)}(k,p)-FL_x^{(\upsilon)}
$$

which yields :

$$
\begin{aligned}
pFLL_x^{(\upsilon)}(k,p)-FL_x^{(\upsilon)} = {} & -8W_O FLL_x^{(\upsilon)}(k,p)+4W_O c_x c_y FLL_y^{(\upsilon)}(k,p)+4W_O c_x c_z FLL_z^{(\upsilon)}(k,p)\big\}\,\mathrm{FLGT}_x^{(\upsilon)}\\
& -e^{-ik\omega_{2x+y+z}}(2S_{y1}^{(\upsilon)})W_O - e^{-ik\omega_{-2x+y+z}}(-2S_{y1}^{(\upsilon)})W_O - e^{-ik\omega_{2x-y+z}}(2S_{y1}^{(\upsilon)})W_O - e^{-ik\omega_{-2x-y+z}}(-2S_{y1}^{(\upsilon)})W_O\\
& -e^{-ik\omega_{2x+y-z}}(2S_{y1}^{(\upsilon)})W_O - e^{-ik\omega_{-2x+y-z}}(-2S_{y1}^{(\upsilon)})W_O - e^{-ik\omega_{2x-y-z}}(2S_{y1}^{(\upsilon)})W_O - e^{-ik\omega_{-2x-y-z}}(-2S_{y1}^{(\upsilon)})W_O\\
& +e^{-ik\omega_{y+z}}\left(8W_O S_{x0}^{(\upsilon)}-2S_{y1}^{(\upsilon)}W_O-(-2S_{y1}^{(\upsilon)})W_O\right)+e^{-ik\omega_{y-z}}\left(8W_O S_{x0}^{(\upsilon)}-2S_{y1}^{(\upsilon)}W_O-(-2S_{y1}^{(\upsilon)})W_O\right)\\
& +e^{-ik\omega_{-y+z}}\left(8W_O S_{x0}^{(\upsilon)}-2S_{y1}^{(\upsilon)}W_O-(-2S_{y1}^{(\upsilon)})W_O\right)+e^{-ik\omega_{-y-z}}\left(8W_O S_{x0}^{(\upsilon)}-2S_{y1}^{(\upsilon)}W_O-(-2S_{y1}^{(\upsilon)})W_O\right)
\end{aligned}
$$

The mirror symmetry plane leads to $S_{x0}^{(\upsilon)}=0$ and to the compensation of the contributions coming from the sinks on planes $x=+1$ and $x=-1$. The last two lines reduce to zero with the final formulation:

$$
pFLL_x^{(\upsilon)}(k,p)-FL_x^{(\upsilon)} = FLGT_x^{(\upsilon)} - 2W_O S_{y1}^{(\upsilon)}\begin{bmatrix} e^{-ik\omega_{2x+y+z}} + e^{-ik\omega_{2x-y+z}} + e^{-ik\omega_{2x+y-z}} + e^{-ik\omega_{2x-y-z}} \\ -e^{-ik\omega_{-2x+y+z}} - e^{-ik\omega_{-2x-y+z}} - + e^{-ik\omega_{-2x+y-z}} - e^{-ik\omega_{-2x-y-z}} \end{bmatrix}
$$

$$
\rightarrow pFLL_x^{(\upsilon)}(k,p)-FLGT_x^{(\upsilon)} = FL_x^{(\upsilon)} + 32ic_x c_y c_z s_x W_O S_{y1}^{(\upsilon)}
$$

Hence the Laplace-Fourier transformed equation becomes:

$$
(p+8W_O)FLL_x^{(\upsilon)} - 4W_O c_x c_y FLL_y^{(\upsilon)} - 4W_O c_x c_z FLL_z^{(\upsilon)} = FL_x^{(\upsilon)} + 32iS_{y1}^{(\upsilon)} W_O c_x c_y c_z s_x \qquad (9)
$$

## 2.2 Transport equations for $L_y^{(\upsilon)}(r,t)$ and $L_z^{(\upsilon)}(r,t)$

The function $L_x^{(\upsilon)}(r,t)$ implies always a displacement of the tracer atom with a non-zero component along Ox. Functions $L_y^{(\upsilon)}(r,t)$ and $L_z^{(\upsilon)}(r,t)$ are different: the divacancy has a first end on the same x plane as that of the tracer atom and a second end in an adjacent x plane. When the first end jumps, the corresponding tracer jump has no component along Ox; but, as a result of this jump having no x-component, the correlation (i.e. the bias between the probabilities of future jumps along +x or –x) does not vanish, unlike the case of the monovacancy mechanism. Indeed, the second end of the divacancy remains the ingredient of a bias between future jumps having a positive or negative component along Ox.

Let us illustrate this point with an example: the divacancy jump from the 'z' site at $\omega_{x+y+2z}$ towards the 'y' site $\omega_{x+z}$ moves the vacancy located on the FCC site $\omega_{2y+2z}$ towards the origin ; the tracer atom is displaced in the reverse direction and the resulting configuration (tracer + divacancy) is identical to the configuration (tracer on the origin + divacancy at the 'y' site $\omega_{x-2y-z}$), apart from a translation of $\omega_{2y+2z}$. This remark implies that a transition (or a fictitious 'jump') of the divacancy from the 'z' site at $\omega_{x+y+2z}$ to the 'y' site at $\omega_{x-2y-z}$ (and vice versa) must be added to account for this configurational change.

The first three lines correspond to the general term for the $L_y(r,t)$ function; the four next to the corrections on sink sites:

$$\frac{dL_y^{(\upsilon)}(r,t)}{dt} = \left.\begin{array}{l} -8W_O L_y^{(\upsilon)}(r,t) \\ +W_O\left[L_x^{(\upsilon)}(r+\omega_{x+y},t)+L_x^{(\upsilon)}(r+\omega_{x-y},t)+L_x^{(\upsilon)}(r+\omega_{-x+y},t)+L_x^{(\upsilon)}(r+\omega_{-x-y},t)\right] \\ +W_O\left[L_z^{(\upsilon)}(r+\omega_{y+z},t)+L_z^{(\upsilon)}(r+\omega_{y-z},t)+L_z^{(\upsilon)}(r+\omega_{-y+z},t)+L_z^{(\upsilon)}(r+\omega_{-y-z},t)\right] \end{array}\right\} GT_y^{(\upsilon)}$$
$$+\delta(r-\omega_{x+z})\left(8W_O L_y^{(\upsilon)}(\omega_{x+z},t)-W_O L_x^{(\upsilon)}(\omega_{y+z},t)-W_O L_x^{(\upsilon)}(\omega_{-y+z},t)-W_O L_z^{(\upsilon)}(\omega_{x+y},t)-W_O L_z^{(\upsilon)}(\omega_{x-y},t)\right)$$
$$+\delta(r-\omega_{x-z})\left(8W_O L_y^{(\upsilon)}(\omega_{x-z},t)-W_O L_x^{(\upsilon)}(\omega_{y-z},t)-W_O L_x^{(\upsilon)}(\omega_{-y-z},t)-W_O L_z^{(\upsilon)}(\omega_{x+y},t)-W_O L_z^{(\upsilon)}(\omega_{x-y},t)\right)$$
$$+\delta(r-\omega_{-x+z})\left(8W_O L_y^{(\upsilon)}(\omega_{-x+z},t)-W_O L_x^{(\upsilon)}(\omega_{y+z},t)-W_O L_x^{(\upsilon)}(\omega_{-y+z},t)-W_O L_z^{(\upsilon)}(\omega_{-x+y},t)-W_O L_z^{(\upsilon)}(\omega_{-x-y},t)\right)$$
$$+\delta(r-\omega_{-x-z})\left(8W_O L_y^{(\upsilon)}(\omega_{-x-z},t)-W_O L_x^{(\upsilon)}(\omega_{y-z},t)-W_O L_x^{(\upsilon)}(\omega_{-y-z},t)-W_O L_z^{(\upsilon)}(\omega_{-x+y},t)-W_O L_z^{(\upsilon)}(\omega_{-x-y},t)\right)$$

At last eight lines corresponding to the four 'y' sites in plane $x=+1$ and four 'y' sites in plane $x=-1$ which are first neighbours of sinks must then be added. The terms corresponding to the extra transitions are written in red below:

$$+\delta(r-\omega_{x+2y+z})\left[-W_O L_x^{(\upsilon)}(\omega_{y+z},t)-W_O L_z^{(\upsilon)}(\omega_{x+y},t) \quad +W_O L_z^{(\upsilon)}(\omega_{x-y-2z},t)\right]$$
$$+\delta(r-\omega_{x+2y-z})\left[-W_O L_x^{(\upsilon)}(\omega_{y-z},t)-W_O L_z^{(\upsilon)}(\omega_{x+y},t) \quad +W_O L_z^{(\upsilon)}(\omega_{x-y+2z},t)\right]$$
$$+\delta(r-\omega_{x-2y+z})\left[-W_O L_x^{(\upsilon)}(\omega_{-y+z},t)-W_O L_z^{(\upsilon)}(\omega_{x+y},t) \quad +W_O L_z^{(\upsilon)}(\omega_{x+y-2z},t)\right]$$
$$+\delta(r-\omega_{x-2y-z})\left[-W_O L_x^{(\upsilon)}(\omega_{-y-z},t)-W_O L_z^{(\upsilon)}(\omega_{x-y},t) \quad +W_O L_z^{(\upsilon)}(\omega_{x+y+2z},t)\right]$$

$$+\delta(r-\omega_{-x+2y+z})\left[-W_O L_x^{(\upsilon)}(\omega_{y+z},t)-W_O L_z^{(\upsilon)}(\omega_{-x+y},t)\ +W_O L_z^{(\upsilon)}(\omega_{-x-y-2z},t)\right]$$
$$+\delta(r-\omega_{-x+2y-z})\left[-W_O L_x^{(\upsilon)}(\omega_{y-z},t)-W_O L_z^{(\upsilon)}(\omega_{-x+y},t)\ +W_O L_z^{(\upsilon)}(\omega_{-x-y+2z},t)\right]$$
$$+\delta(r-\omega_{-x-2y+z})\left[-W_O L_x^{(\upsilon)}(\omega_{-y+z},t)-W_O L_z^{(\upsilon)}(\omega_{-x-y},t)+W_O L_z^{(\upsilon)}(\omega_{-x+y-2z},t)\right]$$
$$+\delta(r-\omega_{-x-2y-z})\left[-W_O L_x^{(\upsilon)}(\omega_{-y-z},t)-W_O L_z^{(\upsilon)}(\omega_{-x-y},t)+W_O L_z^{(\upsilon)}(\omega_{-x+y+2z},t)\right]$$

Fourier-Laplace transforms lead to the final equation for $FLL_y^{(\upsilon)}$ :

$$-4W_O c_y c_x FLL_x^{(\upsilon)}+(p+8W_O)FLL_y^{(\upsilon)}-4W_O c_y c_z FLL_z^{(\upsilon)}=FL_y^{(\upsilon)}+8iW_O S_{y1}^{(\upsilon)}c_z s_x(c_{2y}-3)-8iW_O LL_{y2}^{(\upsilon)}c_z s_x c_{2y}\,. \quad (10)$$

Thanks to the four-fold symmetry around the Ox axis, the equation for $FLL_z^{(\upsilon)}$ is obtained through a mere exchange of 'y' and 'z' :

$$-4W_O c_z c_x FLL_x^{(\upsilon)}-4W_O c_y c_z FLL_y^{(\upsilon)}+(p+8W_O)FLL_z^{(\upsilon)}=FL_z^{(\upsilon)}+8iW_O S_{z1}^{(\upsilon)}c_y s_x(c_{2z}-3)-8iW_O LL_{z2}^{(\upsilon)}c_y s_x c_{2z}\,. \quad (11)$$

## 3. Calculation of return probabilities

The system of Eqs. (9-11) is solved to get the explicit expression of the functions $L_x^{(\upsilon)}(r,t),\ L_y^{(\upsilon)}(r,t),\ L_z^{(\upsilon)}(r,t)$ and determine the unknows $S_{y1}^{(\upsilon)}, LL_{y2}^{(\upsilon)}$.

$$M\begin{pmatrix} FLL_x^{(\upsilon)} \\ FLL_y^{(\upsilon)} \\ FLL_z^{(\upsilon)} \end{pmatrix}=\begin{pmatrix} FL_x^{(\upsilon)}+B_x S_{y1}^{(\upsilon)}+C_x LL_{y2}^{(\upsilon)} \\ FL_y^{(\upsilon)}+B_y S_{y1}^{(\upsilon)}+C_y LL_{y2}^{(\upsilon)} \\ FL_z^{(\upsilon)}+B_z S_{y1}^{(\upsilon)}+C_z LL_{y2}^{(\upsilon)} \end{pmatrix} \text{ with } M=\begin{pmatrix} p+8W_O & -4W_O c_x c_y & -4W_O c_x c_z \\ -4W_O c_y c_x & p+8W_O & -4W_O c_y c_z \\ -4W_O c_z c_x & -4W_O c_z c_y & p+8W_O \end{pmatrix},$$

and

$$\begin{aligned} B_x &= 32iW_O c_x c_y c_z s_x & C_x &= 0 \\ B_y &= 8iW_O c_z s_x(c_{2y}-3) & C_y &= -8iW_O c_z s_x c_{2y}\,. \\ B_z &= 8iW_O c_y s_x(c_{2z}-3) & C_z &= -8iW_O c_y s_x c_{2z} \end{aligned} \quad (12)$$

The general solution

$$\begin{pmatrix} FLL_x^{(\upsilon)} \\ FLL_y^{(\upsilon)} \\ FLL_z^{(\upsilon)} \end{pmatrix}=M^{-1}\begin{pmatrix} FL_x^{(\upsilon)}+B_x S_{y1}^{(\upsilon)}+C_x LL_{y2}^{(\upsilon)} \\ FL_y^{(\upsilon)}+B_y S_{y1}^{(\upsilon)}+C_y LL_{y2}^{(\upsilon)} \\ FL_z^{(\upsilon)}+B_z S_{y1}^{(\upsilon)}+C_z LL_{y2}^{(\upsilon)} \end{pmatrix}$$

requires the expressions of the determinant $\Delta$ and cofactors $D_{ij}$ ; they are given in Appendix A. First order terms with respect to $p$ must be kept all along.

The inverse Fourier transform is used to get the time integral of the looked for probability on any desired site 'r' of type 'c':

$$LL_c^{(\upsilon)}(r,p) = \frac{1}{V_{ZB}} \int_{V_{ZB}} FLL_c^{(\upsilon)}(k,p) e^{+ikr} d_3k \quad (13)$$

where the superscript $(\upsilon)$ keeps track of the initial condition.

### 3.1 Determining $pS_{y1}^{(\upsilon)}$ et $W_O LL_{y2}^{(\upsilon)}$

The system must first be solved for $FLL_y^{(\upsilon)}$ in order to obtain two equations for the two unknowns $S_{y1}^{(\upsilon)}$ et $LL_{y2}^{(\upsilon)}$ :

$$FLL_y^{(\upsilon)} = \frac{N_y^{(\upsilon)}}{\Delta} \qquad \mathrm{N}_y^{(\upsilon)} = \begin{pmatrix} D_{12} & D_{22} & D_{32} \end{pmatrix} \begin{pmatrix} FL_x^{(\upsilon)} + B_x S_{y1}^{(\upsilon)} + C_x LL_{y2}^{(\upsilon)} \\ FL_y^{(\upsilon)} + B_y S_{y1}^{(\upsilon)} + C_y LL_{y2}^{(\upsilon)} \\ FL_z^{(\upsilon)} + B_z S_{y1}^{(\upsilon)} + C_z LL_{y2}^{(\upsilon)} \end{pmatrix}$$

The numerator can be recast into the form $N_y^{(\upsilon)} = N_{y0}^{(\upsilon)} + N_{y1} S_{y1}^{(\upsilon)} + N_{y2} LL_{y2}^{(\upsilon)}$, where the expressions of $N_{y0}^{(\upsilon)}, N_{y1}, N_{y2}$ are established in Appendix B.
Instead of using the value on the particular site $\omega_{x+z}$ only, we perform an average on the four sink sites $\omega_{x+z}, \omega_{x-z} \omega_{-x+z} \omega_{-x-z}$, since the values are equal in magnitude (while taking account of the sign change induces by the mirror symmetry plane):

$$S_{y1}^{(\upsilon)} = \int FLL_y^{(\upsilon)} \frac{e^{+ik\omega_{x+z}} + e^{+ik\omega_{x-z}} - e^{+ik\omega_{-x+z}} - e^{+ik\omega_{-x-z}}}{4} d_3k = i \int FLL_y^{(\upsilon)} s_x c_z d_3k$$
$$= i \int \frac{\left\{ N_{y0}^{(\upsilon)} + N_{y1} S_{y1}^{(\upsilon)} + N_{y2} LL_{y2}^{(\upsilon)} \right\}}{\Delta} s_x c_z d_3k$$

Hence the first equation:

$$S_{y1}^{(\upsilon)} \left\{ 1 - i \int \frac{N_{y1}}{\Delta} s_x c_z d_3k \right\} - i LL_{y2}^{(\upsilon)} \int \frac{N_{y2}}{\Delta} s_x c_z d_3k = i \int \frac{N_{y0}^{(\upsilon)}}{\Delta} s_x c_z d_3k$$

It is shown in Appendix B that the coefficient of $S_{y1}^{(\upsilon)}$ can be recast into the form $pI_{11}/W_O$, and that the right-hand side can be recast into the form $I_{13}^{(\upsilon)}/W_O$. In this first equation connecting the finite quantities $pS_{y1}^{(\upsilon)}$ and $W_O LL_{y2}^{(\upsilon)}$, the integrals $I_{11}, I_{12}, I_{13}^{(\upsilon)}$ are lattice integrals, the expression of which is given in Appendix B:

$$\rightarrow I_{11} pS_{y1}^{(\upsilon)} + I_{12} W_O LL_{y2}^{(\upsilon)} = I_{13}^{(\upsilon)} \quad (14)$$

In the same way, we evaluate $LL_{y2}^{(\upsilon)}$ through an average on eight sites:

$$LL_{y2}^{(\upsilon)} = \int \frac{N_y^{(\upsilon)}}{\Delta} \frac{\begin{Bmatrix} e^{+ik\omega_{x+2y+z}} + e^{+ik\omega_{x+2y-z}} + e^{+ik\omega_{x-2y+z}} + e^{+ik\omega_{x-2y-z}} \\ -e^{+ik\omega_{-x+2y+z}} - e^{+ik\omega_{-x+2y-z}} - e^{+ik\omega_{-x-2y+z}} - e^{+ik\omega_{-x-2y-z}} \end{Bmatrix}}{8} d_3k$$

$$= i\int \frac{N_y^{(\upsilon)}}{\Delta} c_{2y}c_z s_x d_3k$$

Hence the second equation:

$$-iS_{y1}^{(\upsilon)} \int \frac{N_{y1}}{\Delta} c_{2y}c_z s_x d_3k + LL_{y2}^{(\upsilon)} \left\{ 1 - i\int \frac{N_{y2}}{\Delta} c_{2y}c_z s_x d_3k \right\} = i\int \frac{N_{y0}^{(\upsilon)}}{\Delta} c_{2y}c_z s_x d_3k$$

In the same way as above, Appendix B shows that the coefficient of $S_{y1}^{(\upsilon)}$ and the right-hand side are of the form $pI_{21}/W_O$ and $I_{23}^{(\upsilon)}/W_O$ respectively. The second equation can be recast into the form:

$$\rightarrow pI_{21}S_{y1}^{(\upsilon)} + I_{22}W_O LL_{y2}^{(\upsilon)} = I_{23}^{(\upsilon)} \quad (15)$$

The solution of the system is immediate and is given by:

$$pS_{y1}^{(\upsilon)} = \frac{I_{13}^{(\upsilon)}I_{22} - I_{23}^{(\upsilon)}I_{12}}{I_{11}I_{22} - I_{21}I_{12}} \qquad W_O LL_{y2}^{(\upsilon)} = \frac{I_{23}^{(\upsilon)}I_{11} - I_{13}^{(\upsilon)}I_{21}}{I_{11}I_{22} - I_{21}I_{12}} \quad (16)$$

As expected, the two unknowns being time integrals have the physical dimension of a time; the first one is proportional to 1/p; the second one is proportional to $1/W_O$. The formal expressions of integrals $I_{11}, I_{22}, I_{21}, I_{12}, I_{13}^{(\upsilon)}, I_{23}^{(\upsilon)}$ are given in Eqs. (B1-B8); their numerical values are gathered in Table B1. It is also shown that they are not independent but rather linked by three relations established in Eqs. (B9-B11).

### 3.2 Calculation of the total return probabilities

The calculation of the correlation factor implies the total return probability of the divacancy onto the tracer to let it perform a jump with non-zero x projection. The type of the second jump can be of any of the two jump types 1 or 2.

#### 3.2.1 Probability of performing a second tracer jump of type 2

The total probability $P_{\Sigma\, y1}^{(\upsilon)}$ of a second jump of type 2 with positive x-component implies the presence of the divacancy on the four 'y' sites $\omega_{x+2y+z}, \omega_{x+2y-z}, \omega_{x-2y+z}, \omega_{x-2y-z}$ in the plane $x = +1$; the contribution of a jump with a negative x-component is already embedded in the antisymmetric initial condition:

$$
\begin{aligned}
P_{\Sigma y1}^{(\upsilon)} &= W_O \int FLL_y^{(\upsilon)} (e^{+ik\omega_{x+2y+z}} + e^{+ik\omega_{x+2y-z}} + e^{+ik\omega_{x-2y+z}} + e^{+ik\omega_{x-2y-z}}) d_3k \\
&= 4W_O \int FLL_y^{(\upsilon)} (c_x + is_x) c_{2y} c_z d_3k \\
&= P_{y1}^{(\upsilon)} + P_{y11} p S_{y1}^{(\upsilon)} + W_O P_{y12} LL_{y2}^{(\upsilon)}
\end{aligned} \tag{17}
$$

As shown in Appendix C, $P_{y1}^{(\upsilon)}$ and $P_{y11}$ are dimensionless while the third coefficient can be recast into the form $W_O P_{y12}$. Their analytical expressions are given in Eqs. (C1-C4) ; their numerical values are gathered in Table C1. It is also shown that $P_{y11}$, $P_{y12}$ are proportional to $I_{21}$, $I_{22}$ respectively.

#### 3.2.2 Probability of performing a second tracer jump of type 1

The total probability $P_{\Sigma x}^{(\upsilon)}$ of a second tracer jump of type 1 with a positive x-component implies the presence of the vacancy on any of the four 'x' sites $\omega_{2x+y+z}, \omega_{2x+y-z}, \omega_{2x-y+z}, \omega_{2x-y-z}$ in the plane $x = +2$ ; the contribution of a jump with negative x-component is embedded in the initial condition. This probability needs the expression of $FLL_x^{(\upsilon)}(k, p)$ :

$$
FLL_x^{(\upsilon)} = \frac{N_x^{(\upsilon)}}{\Delta} \qquad N_x^{(\upsilon)} = (D_{11} \quad D_{21} \quad D_{31}) \begin{pmatrix} FL_x^{(\upsilon)} + B_x S_{y1}^{(\upsilon)} \\ FL_y^{(\upsilon)} + B_y S_{y1}^{(\upsilon)} + C_y LL_{y2}^{(\upsilon)} \\ FL_z^{(\upsilon)} + B_z S_{y1}^{(\upsilon)} + C_z LL_{y2}^{(\upsilon)} \end{pmatrix}
$$

$$
\begin{aligned}
N_x^{(\upsilon)} &= (D_{11} FL_x^{(\upsilon)} + D_{21} FL_y^{(\upsilon)} + D_{31} FL_z^{(\upsilon)}) + (D_{11} B_x + D_{21} B_y + D_{31} B_z) S_{y1}^{(\upsilon)} \\
&\quad + (D_{11} C_x + D_{21} C_y + D_{31} C_z) LL_{y2}^{(\upsilon)}
\end{aligned}
$$

As above it must be Fourier inversed and the average is performed on the four sites:

$$
\begin{aligned}
P_{\Sigma x}^{(\upsilon)} &= 2W_O \int FLL_x^{(\upsilon)} (e^{+ik\omega_{2x+y+z}} + e^{+ik\omega_{2x+y-z}} + e^{+ik\omega_{2x-y+z}} + e^{+ik\omega_{2x-y-z}}) d_3k \\
&= 8W_O \int FLL_x^{(\upsilon)} (c_{2x} + is_{2x}) c_y c_z d_3k \\
&= P_x^{(\upsilon)} + P_{x1} p S_{y1}^{(\upsilon)} + W_O P_{x2} LL_{y2}^{(\upsilon)}
\end{aligned} \tag{18}
$$

As above, the coefficient of $S_{y1}^{(\upsilon)}$ is shown in Appendix D to be equal to $pP_{x1}$. The expressions of the coefficients $P_x^{(\upsilon)}$, $P_{x1}$, $P_{x2}$ are established in Eqs. (D1-D4) and their numerical values gathered in Table D1. It is also shown that $P_{x1}$, $P_{x2}$ are proportionnal to $I_{13}^{(2)}$, $I_{23}^{(2)}$ respectively according to Eqs. (D6-D7).

### 3.3 Expressions of $t_{ij}$ and correlation factor

We calculated above the contribution $P_{\Sigma y1}^{(1)}$ of divacancy jumps starting from any of the four 'y' sites of plane $x = +1$. The contribution $P_{\Sigma z1}^{(1)}$ of the divacancy jumps starting from the four 'z' sites on plane $x = +1$ is of equal magnitude thanks to the four-fold symmetry assigning the same role to 'y' and 'z' variables. Hence the final average products $t_{ij}$ entering the expression of the correlation factor:

$$\begin{aligned} t_{11} &= -P^{(1)}_{\Sigma x} \qquad & t_{12} &= -(P^{(1)}_{\Sigma y1} + P^{(1)}_{\Sigma z1}) = -2P^{(1)}_{\Sigma y1} \\ t_{21} &= -P^{(2)}_{\Sigma x} \qquad & t_{22} &= -(P^{(2)}_{\Sigma y1} + P^{(2)}_{\Sigma z1}) = -2P^{(2)}_{\Sigma y1} \end{aligned} \tag{19}$$

Appendix E establishes the equality $t_{11} = t_{22}$ in Eqs. (E2-E3).
The numerical values of $t_{ij}$'s are gathered in Table E1. The final value of the correlation factor $f_{2V}$ is thus equal to:

$$f_{2V} = 0.45809434 \tag{20}$$

This exact value is in the confidence interval of the two best evaluations known so far.[6-7]

**Acknowledgements**

We thank warmly J. Creuze from LEMHE-ICMMO, Orsay (France), with whom this work was started when he was a M2 student.

**Appendix A: Expressions of determinant $\Delta$ and cofactors $D_{ij}$ at first order in p**

After retaining only the terms at first order in p :

$$\Delta = \begin{vmatrix} p+8W_O & -4W_O c_x c_y & -4W_O c_x c_z \\ -4W_O c_y c_x & p+8W_O & -4W_O c_y c_z \\ -4W_O c_z c_x & -4W_O c_z c_y & p+8W_O \end{vmatrix} \approx \Delta_0 + p\Delta_1 \qquad \text{(A1)}$$

$$\Delta_0 = 128W_O^3\left[4 - c_x^2c_y^2c_z^2 - c_x^2c_y^2 - c_y^2c_z^2 - c_z^2c_x^2\right] = 128W_O^3\Delta_0^{'}$$
$$\Delta_1 = 16W_O^2\left[12 - c_x^2c_y^2 - c_y^2c_z^2 - c_z^2c_x^2\right] = 16W_O^2\Delta_1^{'} \qquad \text{(A2)}$$
$$\Delta = \Delta_0(1 + p\frac{\Delta_1}{\Delta_0}) \rightarrow \Delta = 128W_O^3\Delta_0^{'}\left(1 + p\frac{\Delta_1^{'}}{8W_O\Delta_0^{'}}\right)$$

Alternative expressions of determinant as a function of cofactors $D_{ij}$ :

$$\Delta = (p+8W_O)D_{11} - 4W_O c_y c_x D_{21} - 4W_O c_z c_x D_{31}$$
$$\text{with } \begin{cases} D_{11} \approx 64W_O^2 - 16W_O^2c_y^2c_z^2 + p16W_O \\ D_{21} = 16W_O^2c_xc_yc_z^2 + 32W_O^2c_xc_y + p4W_Oc_xc_y \\ D_{31} = 16W_O^2c_xc_y^2c_z + 32W_O^2c_xc_z + p4W_Oc_xc_z \end{cases} \qquad \text{(A3)}$$

or

$$\Delta = -4W_O c_x c_y D_{12} + (p+8W_O)D_{22} - 4W_O c_y c_z D_{32}$$
$$\text{with } \begin{cases} D_{12} = 16W_O^2c_xc_yc_z^2 + 32W_O^2c_xc_y + p4W_Oc_xc_y \\ D_{22} \approx 64W_O^2 - 16W_O^2c_x^2c_z^2 + p16W_O \\ D_{32} = 16W_O^2c_x^2c_yc_z + 32W_O^2c_yc_z + p4W_Oc_yc_z \end{cases} \qquad \text{(A4)}$$

or

$$\Delta = -4W_O c_x c_z D_{13} - 4W_O c_y c_z D_{23} + (p+8W_O)D_{33}$$
$$\text{with } \begin{cases} D_{13} = 16W_O^2c_xc_y^2c_z + 32W_O^2c_xc_z + p4W_Oc_xc_z \\ D_{23} = 16W_O^2c_x^2c_yc_z + 32W_O^2c_yc_z + p4W_Oc_yc_z \\ D_{33} \approx 64W_O^2 - 16W_O^2c_x^2c_y^2 + p16W_O \end{cases} \qquad \text{(A5)}$$

Symmetry implies : $D_{12} = D_{21}$ $\quad D_{23} = D_{32}$ $\quad D_{31} = D_{13}$

## Appendix B: Integral expressions of the coefficients $I_{ij}^{(\upsilon)}$ in the equations yielding $pS_{y1}^{(\upsilon)}$ and $LL_{y2}^{(\upsilon)}$

$D_{ij}$ stand for the cofactors calculated in Appendix A (Eqs. A3-A5).
In all what follows the normalizing factor $1/V_{ZB}$ of the integrals is omitted for simplicity.

**B1. Calculation of** $\frac{pI_{11}}{W_O}=1-i\int\frac{D_{12}B_x+D_{22}B_y+D_{32}B_z}{\Delta}s_xc_zd_3k=1+\int\frac{N_{11}}{\Delta}d_3k$

Numerator $N_{11}$:

$$B_x=32iW_Oc_xc_yc_zs_x \qquad B_y=8iW_Oc_zs_x(c_{2y}-3) \qquad B_z=8iW_Oc_ys_x(c_{2z}-3)$$

$$N_{y1}=D_{12}B_x+D_{22}B_y+D_{32}B_z=-512iW_O^3c_zs_x\Delta_O'+64ipW_O^2c_zs_x\left\{2c_x^2c_y^2+c_y^2c_z^2+2c_y^2-8\right\}$$

Multiplying by the integrating factor :

$$N_{11}=-is_xc_zN_{y1}=-512W_O^3c_z^2s_x^2\Delta_O'\left\{1-p\frac{\left[2c_x^2c_y^2+c_y^2c_z^2+2c_y^2-8\right]}{8W_O\Delta_O'}\right\}$$

Hence, retaining only first order term with respect to $p$ yields:

$$I_{11}=1+\int\frac{N_{11}}{\Delta}d_3k=1-\int\frac{4c_z^2s_x^2\left\{1-p\frac{\left[2c_x^2c_y^2+c_y^2c_z^2+2c_y^2-8\right]}{8W_O\Delta_O'}\right\}}{\left(1+p\frac{\Delta_1'}{8W_O\Delta_0'}\right)}d_3k$$

$$=1-\int 4c_z^2s_x^2\left\{1-p\frac{\left[2c_x^2c_y^2+c_y^2c_z^2+2c_y^2-8\right]+\Delta_1'}{8W_O\Delta_O'}\right\}d_3k$$

$$=1-\int 4c_z^2s_x^2d_3k+p\int c_z^2s_x^2\frac{c_x^2c_y^2-c_x^2c_z^2+2c_y^2+4}{2W_O\Delta_O'}d_3k$$

The first integral is equal to 1, hence:

$$\frac{pI_{11}}{W_O}=p\int\frac{c_z^2s_x^2\left\{c_x^2c_y^2-c_x^2c_z^2+2c_y^2+4\right\}}{2W_O\Delta_O'}d_3k \tag{B1}$$

The integral $S_{y1}^{(\upsilon)}$ for the sink sites at $\omega_{x\pm y}$ and $\omega_{x\pm z}$ appears in the first equation with the variable ‘p’ as a multiplying factor as mentioned in the main section.

**B2. Calculation of** $I_{12} = -i\int \frac{D_{12}C_x + D_{22}C_y + D_{32}C_z}{\Delta} s_x c_z d_3k = \int \frac{N_{12}}{\Delta} d_3k$

Expression of $N_{12}$:

$$C_x = 0 \qquad C_y = -8iW_O c_z s_x c_{2y} \qquad C_z = -8iW_O c_y s_x c_{2z}$$

$$N_{y2} = D_{12}C_x + D_{22}C_y + D_{32}C_z$$

$$= -128iW_O^3 c_z s_x \left\{-c_x^2c_y^2 + 4c_y^2c_z^2 + c_x^2c_z^2 + 6c_y^2 - 4\right\} - p32iW_O^2 c_z s_x \left\{4c_{2y} + c_y^2 c_{2z}\right\}$$

The first term which is independent of 'p' gives a non-vanishing contribution to the integral; as a consequence, the first order term in $p$ can be dropped:

$$N_{12} = -is_x c_z N_{y2}$$

$$= -128W_O^3 c_z^2 s_x^2 \left[-c_x^2c_y^2 + 4c_y^2c_z^2 + c_x^2c_z^2 + 6c_y^2 - 4\right]$$

$$I_{12} = -\int \frac{128W_O^3 c_z^2 s_x^2 \left[-c_x^2c_y^2 + 4c_y^2c_z^2 + c_x^2c_z^2 + 6c_y^2 - 4\right]}{128W_O^3 \Delta_O'} d_3k$$

Hence :

$$I_{12} = -\int \frac{c_z^2 s_x^2 \left[-c_x^2c_y^2 + 4c_y^2c_z^2 + c_x^2c_z^2 + 6c_y^2 - 4\right]}{\Delta_O'} d_3k \qquad \text{(B2)}$$

**B3. Calculation of** $\frac{I_{13}^{(\upsilon)}}{W_O} = i\int \frac{D_{12}FL_x^{(\upsilon)} + D_{22}FL_y^{(\upsilon)} + D_{32}FL_z^{(\upsilon)}}{\Delta} s_x c_z d_3k = \int \frac{N_{13}^{(\upsilon)}}{\Delta} d_3k$

Numerator $N_{13}^{(1)}$ and $N_{13}^{(2)}$:

(first order terms in $p$ can be dropped in $D_{ij}$ ).

For CI1 : $FL_x^{(1)} = 0 \qquad FL_y^{(1)} = -ic_{2y}c_z s_x \qquad FL_z^{(1)} = -ic_y c_{2z} s_x$

$$N_{y0}^{(1)} = D_{12}FL_x^{(1)} + D_{22}FL_y^{(1)} + D_{32}FL_z^{(1)} = -16iW_O^2 c_z s_x \left\{c_{2y}\left[4 - c_x^2c_z^2\right] + c_{2z}\left[c_x^2c_y^2 + 2c_y^2\right]\right\}$$

$$N_{13}^{(1)} = iN_{y0}^{(1)} s_x c_z = 16W_O^2 c_z^2 s_x^2 \left\{c_{2y}\left[4 - c_x^2c_z^2\right] + c_{2z}\left[c_x^2c_y^2 + 2c_y^2\right]\right\}$$

Hence :

$$\frac{I_{13}^{(1)}}{W_O} = \int \frac{c_z^2 s_x^2 \left\{c_x^2c_z^2 + 4c_y^2c_z^2 - c_x^2c_y^2 + 6c_y^2 - 4\right\}}{8W_O \Delta_O'} d_3k \qquad \text{(B3)}$$

For CI2 : $FL_x^{(2)} = -4ic_x c_y c_z s_x \qquad FL_y^{(2)} = 0 \qquad FL_z^{(2)} = 0$

$$N_{y0}^{(2)} = D_{12}FL_x^{(2)} = -64iW_O^2c_x^2c_y^2c_zs_x\left\{c_z^2+2\right\}$$
$$N_{13}^{(2)} = iN_{y0}^{(2)}s_xc_z = 64W_O^2c_x^2c_y^2c_z^2s_x^2\left\{c_z^2+2\right\}$$

Hence :

$$\frac{I_{13}^{(2)}}{W_O} = \int \frac{c_x^2c_y^2c_z^2s_x^2\left\{c_z^2+2\right\}}{2W_O\Delta_O^{'}}d_3k \tag{B4}$$

**Calculation of** $\frac{pI_{21}}{W_O} = -i\int \frac{D_{12}B_x + D_{22}B_y + D_{32}B_z}{\Delta}c_{2y}c_zs_xd_3k = \int \frac{N_{21}}{\Delta}d_3k$

Numerator $N_{21} = c_{2y}N_{11}$

Hence : $\frac{pI_{21}}{W_O} = -\int 4c_z^2s_x^2c_{2y}d_3k + p\int c_z^2s_x^2c_{2y}\frac{c_x^2c_y^2 - c_x^2c_z^2 + 2c_y^2 + 4}{2W_O\Delta_O^{'}}d_3k$

The first integral is equal to zero, hence:

$$\frac{pI_{21}}{W_O} = p\int \frac{c_z^2s_x^2c_{2y}\left[c_x^2c_y^2 - c_x^2c_z^2 + 2c_y^2 + 4\right]}{2W_O\Delta_O^{'}}d_3k \tag{B5}$$

**Calculation of** $I_{22} = 1 - i\int \frac{D_{12}C_x + D_{22}C_y + D_{32}C_z}{\Delta}c_{2y}c_zs_xd_3k = 1 - \int \frac{N_{22}}{\Delta}d_3k$

Numerator $N_{y2} = c_{2y}N_{12}$

Hence :

$$I_{22} = 1 - \int \frac{c_z^2s_x^2c_{2y}\left\{-c_x^2c_y^2 + 4c_y^2c_z^2 + c_x^2c_z^2 + 6c_y^2 - 4\right\}}{\Delta_O^{'}}d_3k \tag{B6}$$

**Calculation of** $\frac{I_{23}^{(\upsilon)}}{W_O} = \int \frac{N_{23}^{(\upsilon)}}{\Delta}d_3k$

Numerator $N_{23}^{(1)}$ for Cl1 :

$$N_{23}^{(1)} = c_{2y}N_{13}^{(1)} = 16W_O^2c_z^2c_{2y}s_x^2\left\{c_x^2c_z^2 + 4c_y^2c_z^2 - c_x^2c_y^2 + 6c_y^2 - 4\right\}$$
$$\frac{I_{23}^{(1)}}{W_O} = \int \frac{16W_O^2c_z^2c_{2y}s_x^2\left\{c_x^2c_z^2 + 4c_y^2c_z^2 - c_x^2c_y^2 + 6c_y^2 - 4\right\}}{128W_O^3\Delta_O^{'}}d_3k$$

Hence :

$$\frac{I_{23}^{(1)}}{W_O} = \int \frac{c_z^2 c_{2y} s_x^2 \left\{c_x^2 c_z^2 + 4c_y^2 c_z^2 - c_x^2 c_y^2 + 6c_y^2 - 4\right\}}{8W_O \Delta_O^{'}} d_3k \tag{B7}$$

Numerator $N_{23}^{(2)}$ for CI2:

$$N_{23}^{(2)} = c_{2y} N_{13}^{(2)} = 64W_O^2 c_x^2 c_y^2 c_z^2 c_{2y} s_x^2 \left\{c_z^2 + 2\right\}$$

$$\frac{I_{23}^{(2)}}{W_O} = \int \frac{64W_O^2 c_x^2 c_y^2 c_z^2 c_{2y} s_x^2 \left\{c_z^2 + 2\right\}}{128W_O^3 \Delta_O^{'}} d_3k$$

Hence :

$$\frac{I_{23}^{(2)}}{W_O} = \int \frac{c_x^2 c_y^2 c_z^2 c_{2y} s_x^2 \left\{c_z^2 + 2\right\}}{2W_O \Delta_O^{'}} d_3k \tag{B8}$$

## Relationships between integrals

All the lattice integrals given above are not independent. We derive several useful relationships to be used later on. In order to recast the expression of an integral into an alternative formulation, we take into account the fact that the variables ‘y’ and ‘z’ play the same role and can be interchanged at will.
For instance :

$$I_{13}^{(2)} = \int \frac{c_x^2 c_y^2 c_z^2 s_x^2 \left\{c_z^2 + 2\right\}}{2\Delta_O^{'}} d_3k = \int \frac{c_x^2 c_y^2 c_z^2 s_x^2 \left\{c_y^2 + 2\right\}}{2\Delta_O^{'}} d_3k = \int \frac{c_x^2 c_y^2 c_z^2 s_x^2 \left\{c_y^2 + c_z^2 + 4\right\}}{4\Delta_O^{'}} d_3k$$

### Relationship between $I_{12}$ and $I_{13}^{(1)}$

An obvious result after comparing Eqs (B2-B3):

$$I_{12} = -8I_{13}^{(1)} \tag{B9}$$

### Relationship between $I_{21}$ and $I_{12}$

Starting from Eq. B5 and expanding $c_{2y}$ as a function of $c_y$ gives two parts :

$$2I_{21} = \int \frac{2c_y^2 c_z^2 s_x^2 \left[c_x^2 c_y^2 - c_x^2 c_z^2 + 2c_y^2 + 4\right]}{\Delta_O^{'}} d_3k - \int \frac{c_z^2 s_x^2 \left[c_x^2 c_y^2 - c_x^2 c_z^2 + 2c_y^2 + 4\right]}{\Delta_O^{'}} d_3k$$

The first part is written as half the sum of two equivalent contributions after exchanging ‘y’ and ‘z’:

$$2I_{21}=\int\frac{c_y^2c_z^2s_x^2\left[c_x^2c_y^2-c_x^2c_z^2+2c_y^2+4\right]}{\Delta_O^{'}}d_3k+\int\frac{c_y^2c_z^2s_x^2\left[c_x^2c_z^2-c_x^2c_y^2+2c_z^2+4\right]}{\Delta_O^{'}}$$
$$-\int\frac{c_z^2s_x^2\left[c_x^2c_y^2-c_x^2c_z^2+2c_y^2+4\right]}{\Delta_O^{'}}d_3k$$

Summing the two first contributions gives :

$$2I_{21}=\int\frac{4c_y^2c_z^2s_x^2\left[c_y^2+2\right]}{\Delta_O^{'}}-\int\frac{c_z^2s_x^2\left[c_x^2c_y^2-c_x^2c_z^2+2c_y^2+4\right]}{\Delta_O^{'}}d_3k\,,$$

and finally:

$$2I_{21}=\int\frac{c_z^2s_x^2\left[-c_x^2c_y^2+4c_y^4+c_x^2c_z^2+6c_y^2-4\right]}{\Delta_O^{'}}$$

Adding Eq. B2 yields:

$$2I_{21}+I_{12}=\int\frac{4c_y^2c_z^2s_x^2\left[c_y^2-c_z^2\right]}{\Delta_O^{'}}=\int\frac{4c_y^2c_z^2s_x^2\left[c_z^2-c_y^2\right]}{\Delta_O^{'}}$$

The right-hand side is equal to zero thanks to the symmetric roles played by ‘y’ and ‘z’, hence :

$$2I_{21}=-I_{12} \tag{B10}$$

**Relationship between $I_{22}$ and $I_{21}$**

Starting from Eqs. (B5-B6):

$$2I_{21}-I_{22}+1=\int\frac{4c_{2y}c_y^2c_z^2s_x^2\left\{c_z^2+2\right\}}{\Delta_O^{'}}d_3k \tag{B11}$$

**Relationship between $I_{22}$ and $I_{23}^{(1)}$**

Comparing Eqs. (B6-B7) yields:

$$1-I_{22}=8I_{23}^{(1)} \tag{B12}$$

The independent integrals entering the coefficients of the linear system yielding the two unknows $pS_{y1}^{(\upsilon)}$ and $W_OLL_{y2}^{(\upsilon)}$ can be calculated numerically with a superposition of three Gaussian quadratures along the three axes with 512 nodes on the interval $\left[0,+\pi\right]$. They are gathered below in Table B1 together with the values of the unknowns for the two initial conditions.

| | | CI1 | CI2 |
|---|---|---|---|
| $I_{11} = 2.136053180D-01$ | $I_{12} = -1.036809774\mathrm{D}-01$ | $I_{13}^{(1)} = -I_{12}/8$ | $I_{13}^{(2)} = 2.224386633\mathrm{D}-02$ |
| $I_{21} = -I_{12}/2$ | $I_{22} = 7.992024140\mathrm{D}-01$ | $I_{23}^{(1)} = (1-I_{22})/8$ | $I_{23}^{(2)} = 1.378066830\mathrm{D}-02$ |
| | | $pS_{y1}^{(1)} = 7.359994075\mathrm{D}-02$ | $pS_{y1}^{(2)} = 1.090708175\mathrm{D}-01$ |
| | | $W_O LL_{y2}^{(1)} = 2.663185317\mathrm{D}-02$ | $W_O LL_{y2}^{(2)} = 1.016811720\mathrm{D}-02$ |

*Table B1. Numerical values of integrals. The solutions corresponding to CI1 and CI2 are displayed in lines and columns 3 and 4.*

## Appendix C: Expression of the coefficients required for the probability of a second tracer jump of type 2

$$P_{\Sigma y1}^{(\upsilon)} = P_{y1}^{(\upsilon)} + P_{y11} p S_{y1}^{(\upsilon)} + W_O P_{y12} LL_{y2}^{(\upsilon)}$$

### First coefficient $P_{y1}^{(1)}$ and $P_{y1}^{(2)}$

$$P_{y1}^{(\upsilon)} = 4W_O \int \frac{D_{12}FL_x^{(\upsilon)} + D_{22}FL_y^{(\upsilon)} + D_{32}FL_z^{(\upsilon)}}{\Delta}(c_x + is_x)c_{2y}c_z d_3k = \int \frac{Num_{y1}^{(\upsilon)}}{\Delta} d_3k$$

For CI1 :

$$D_{12}FL_x^{(1)} + D_{22}FL_y^{(1)} + D_{32}FL_z^{(1)} = N_{y0}^{(1)} = -16iW_O^2 c_z s_x \left\{c_{2y}\left[4 - c_x^2c_z^2\right] + c_{2z}\left[c_x^2c_y^2 + 2c_y^2\right]\right\}$$

$$Num_{y1}^{(1)} = 4W_O N_{y0}^{(1)}(c_x + is_x)c_{2y}c_z = 64W_O^3 c_{2y}c_z^2 s_x^2 \left\{c_x^2c_z^2 + 4c_y^2c_z^2 - c_x^2c_y^2 + 6c_y^2 - 4\right\}$$

Hence :

$$P_{y1}^{(1)} = \int \frac{c_{2y}c_z^2 s_x^2 \left\{c_x^2c_z^2 + 4c_y^2c_z^2 - c_x^2c_y^2 + 6c_y^2 - 4\right\}}{2\Delta_O^{'}} d_3k \qquad \text{(C1)}$$

For CI2 :

$$D_{12}FL_x^{(2)} + D_{22}FL_y^{(2)} + D_{32}FL_z^{(2)} = N_{y0}^{(2)}$$

$$Num_{y1}^{(2)} = 4W_O N_{y0}^{(2)}(c_x + is_x)c_{2y}c_z = 256W_O^3 c_x^2c_y^2c_z^2c_{2y}s_x^2 \left\{c_z^2 + 2\right\}$$

Hence :

$$P_{y1}^{(2)} = \int \frac{2c_x^2c_y^2c_z^2c_{2y}s_x^2 \left\{c_z^2 + 2\right\}}{\Delta_O^{'}} d_3k \qquad \text{(C2)}$$

### Second coefficient

$$P_{y11} = 4W_O \int \frac{\left(D_{12}B_x + D_{22}B_y + D_{32}B_z\right)}{\Delta}(c_x + is_x)c_{2y}c_z d_3k = \int \frac{Num_{y11}}{\Delta} d_3k$$

Retaining the first order term in p and keeping only the real part :

$$\begin{aligned} Num_{y11} &= 4W_O(c_x + is_x)c_{2y}c_z N_{y1} \\ &= 2048W_O^4 c_z^2 s_x^2 c_{2y}\Delta_O^{'}\left\{1 - p\frac{\left[2c_x^2c_y^2 + c_y^2c_z^2 + 2c_y^2 - 8\right]}{8W_O\Delta_O^{'}}\right\} \end{aligned}$$

Hence :

$$P_{y11} = \int 16W_O c_z^2 s_x^2 c_{2y} d_3k - \int 2c_z^2 s_x^2 c_{2y} \left\{ \frac{\left[2c_x^2 c_y^2 + c_y^2 c_z^2 + 2c_y^2 - 8\right] + \Delta_1^{'}}{\Delta_O^{'}} \right\} d_3k$$

The first integral is equal to zero, which yields :

$$pP_{y11} = -p\int \frac{2c_{2y}c_z^2 s_x^2 \left[c_x^2 c_y^2 - c_x^2 c_z^2 + 2c_y^2 + 4\right]}{\Delta_O^{'}} d_3k \tag{C3}$$

**Third coefficient**

$$W_O P_{y12} = 4W_O \int \frac{\left(D_{12}C_x + D_{22}C_y + D_{32}C_z\right)}{\Delta}(c_x + is_x)c_{2y}c_z d_3k = \int \frac{Num_{y12}}{\Delta} d_3k$$

Only terms not depending on $p$ can be retained:

$$Num_{y12} = 4W_O(c_x + is_x)c_{2y}c_z N_{y2} = 512W_O^4 c_z^2 c_{2y} s_x^2 \left\{-c_x^2 c_y^2 + 4c_y^2 c_z^2 + c_x^2 c_z^2 + 6c_y^2 - 4\right\}$$

Hence :

$$W_O P_{y12} = W_O \int \frac{4c_{2y}c_z^2 s_x^2 \left\{-c_x^2 c_y^2 + 4c_y^2 c_z^2 + c_x^2 c_z^2 + 6c_y^2 - 4\right\}}{\Delta_O^{'}} d_3k \tag{C4}$$

The numerical values of the coefficients are gathered in Table C1.

| | | |
|---|---|---|
| $P_{y1}^{(1)} = 1.003987929\text{D}-01$ | $P_{y11} = -2.073619549\text{D}-01$ | $P_{y12} = 8.031903439\text{D}-01$ |
| $P_{y1}^{(2)} = 5.512267320\text{D}-02$ | | |

*Table C1. Coefficients entering the definition of $P_{\Sigma y1}^{(\upsilon)}$ for the two initial conditions CI1 and CI2.*

**Relationship between integrals**

The new integrals are not independent.

Comparison of Eq. C3-C4 with Eq. B11 yields :

$$P_{y12} - 2P_{y11} = 8I_{21} - 4I_{22} + 4 \tag{C5}$$

**Relationship between $P_{y11}$ and $I_{21}$ :**

$$P_{y11} = -\int \frac{2c_{2y}c_z^2 s_x^2 \left[ c_x^2 c_y^2 - c_x^2 c_z^2 + 2c_y^2 + 4 \right]}{\Delta_O^{'}} d_3k$$

$$4I_{21} = \int \frac{2c_{2y}c_z^2 s_x^2 \left[ c_x^2 c_y^2 - c_x^2 c_z^2 + 2c_y^2 + 4 \right]}{\Delta_O^{'}} d_3k$$

Comparison of Eq. C3 with Eq. B6 yields :

$$P_{y11} = -4I_{21} \tag{C6}$$

At last, using Eqs. C5-C6, we get :

$$P_{y12} = 4(1 - I_{22}) \tag{C7}$$

## Appendix D: Expression of the coefficients entering the probability of a second tracer jump of type 1

$$P_{\Sigma x}^{(\upsilon)} = P_x^{(\upsilon)} + P_{x1}pS_{y1}^{(\upsilon)} + W_O P_{x2} LL_{y2}^{(\upsilon)}$$

### First coefficient

$$P_x^{(\upsilon)} = 8W_O \int \frac{(D_{11}FL_x^{(\upsilon)} + D_{21}FL_y^{(\upsilon)} + D_{31}FL_z^{(\upsilon)})}{\Delta}(c_{2x} + is_{2x})c_y c_z d_3k = \int \frac{Num_x^{(\upsilon)}}{\Delta} d_3k$$

The integral depends on the initial conditions CI1 or CI2 and on the expressions of the cofactors (Eqs. A3-A5). The variable $p$ can be set to zero in all what follows.

For CI1: $FL_x^{(1)} = 0$ $\quad FL_y^{(1)} = -ic_{2y}c_z s_x$ $\quad FL_z^{(1)} = -ic_y c_{2z} s_x$

$$Num_x^{(1)} = 256W_O^3 c_x^2 c_y^2 c_z^2 s_x^2 \left\{ c_{2y}\left[c_z^2 + 2\right] + c_{2z}\left[c_y^2 + 2\right]\right\}$$

Hence:

$$P_x^{(1)} = \int \frac{2c_x^2 c_y^2 c_z^2 s_x^2 \left[4c_y^2 c_z^2 + 3c_y^2 + 3c_z^2 - 4\right]}{\Delta_O^{'}} d_3k \tag{D1}$$

For CI2: $FL_x^{(2)} = -4ic_x c_y c_z s_x$ $\quad FL_y^{(2)} = 0$ $\quad FL_z^{(2)} = 0$

$$Num_x^{(2)} = 8W_O D_{11} FL_x^{(2)}(c_{2x} + is_{2x})c_y c_z = 1024W_O^3\left[4 - c_y^2 c_z^2\right] c_x^2 c_y^2 c_z^2 s_x^2$$

Hence :

$$P_x^{(2)} = \int \frac{8c_x^2 c_y^2 c_z^2 s_x^2 \left[4 - c_y^2 c_z^2\right]}{\Delta_O^{'}} d_3k \tag{D2}$$

### Second coefficient

$$pP_{x1} = 8W_O \int \frac{(D_{11}B_x + D_{21}B_y + D_{31}B_z)}{\Delta}(c_{2x} + is_{2x})c_y c_z d_3k = \int \frac{Num_{x1}}{\Delta} d_3k$$

First order terms in $p$ are retained. In this case, the constant terms cancel each other and only the first order terms proportional to p remain in the product:

$$B_x = 32iW_O c_x c_y c_z s_x \qquad B_y = 8iW_O c_z s_x (c_{2y} - 3) \qquad B_z = 8iW_O c_y s_x (c_{2z} - 3)$$

$$Num_{x1} = -p1024W_O^3 c_x^2 c_y^2 c_z^2 s_x^2 \left\{4 + c_y^2 + c_z^2\right\}$$

Hence :

$$pP_{x1} = -p\int \frac{8c_x^2 c_y^2 c_z^2 s_x^2 \left\{4 + c_y^2 + c_z^2\right\}}{\Delta_O^{'}} d_3k \tag{D3}$$

**Third coefficient**

$$W_O P_{x2} = 8W_O \int \frac{(D_{11}C_x + D_{21}C_y + D_{31}C_z)}{\Delta}(c_{2x} + is_{2x})c_y c_z d_3k = \int \frac{Num_{x2}}{\Delta} d_3k:$$

Only terms not depending on $p$ can be retained:

$$C_x = 0 \qquad C_y = -8iW_O c_z s_x c_{2y} \qquad C_z = -8iW_O c_y s_x c_{2z}$$

$$Num_{x2} = 2048 W_O^4 c_x^2 c_y^2 c_z^2 s_x^2 \left\{4c_y^2 c_z^2 + 3c_y^2 + 3c_z^2 - 4\right\}$$

Hence :

$$W_O P_{x2} = W_O \int \frac{16 c_x^2 c_y^2 c_z^2 s_x^2 \left\{4c_y^2 c_z^2 + 3c_y^2 + 3c_z^2 - 4\right\}}{\Delta_O^{'}} d_3k \qquad \text{(D4)}$$

The numerical values of the integrals are gathered in Table D1 below.

| | | |
|---|---|---|
| $P_x^{(1)} = 1.102453464\text{D}-01$ | $P_{x1} = -7.118037228\text{D}-01$ | $P_{x2} = 8.819627712\text{D}-01$ |
| $P_x^{(2)} = 4.236074457\text{D}-01$ | | |

*Table D1. Coefficients entering the definition of $P_{\Sigma x}^{(\upsilon)}$ for the two initial conditions CI1 and CI2.*

**Relationships between integrals**

**Relationship between $P_x^{(1)}$ and $I_{23}^{(2)}$**

$$P_x^{(1)} = \int \frac{2c_x^2 c_y^2 c_z^2 s_x^2 \left[4c_y^2 c_z^2 + 3c_y^2 + 3c_z^2 - 4\right]}{\Delta_O^{'}} d_3k$$

$$I_{23}^{(2)} = \int \frac{c_{2y} c_x^2 c_y^2 c_z^2 s_x^2 \left\{c_z^2 + 2\right\}}{2\Delta_O^{'}} d_3k = \int \frac{(2c_y^2 - 1)c_x^2 c_y^2 c_z^2 s_x^2 \left\{c_z^2 + 2\right\}}{2\Delta_O^{'}} d_3k$$

$$= \int \frac{c_x^2 c_y^2 c_z^2 s_x^2 \left\{\left(2c_y^2 c_z^2 - c_z^2 + 4c_y^2 - 2\right) + \left(2c_y^2 c_z^2 - c_y^2 + 4c_z^2 - 2\right)\right\}}{4\Delta_O^{'}} d_3k$$

$$= \int \frac{c_x^2 c_y^2 c_z^2 s_x^2 \left\{4c_y^2 c_z^2 + 3c_z^2 + 3c_y^2 - 4\right\}}{4\Delta_O^{'}} d_3k$$

Hence :

$$P_x^{(1)} = 8 I_{23}^{(2)} \qquad \text{(D5)}$$

**Relationship between $P_{x1}$ and $I_{13}^{(2)}$**

$$P_{x1} = -\int \frac{8c_x^2 c_y^2 c_z^2 s_x^2 \left\{4 + c_y^2 + c_z^2\right\}}{\Delta_O^{'}} d_3k = -\int \frac{16c_x^2 c_y^2 c_z^2 s_x^2 \left\{2 + c_z^2\right\}}{\Delta_O^{'}} d_3k$$

$$I_{13}^{(2)} = \int \frac{c_x^2 c_y^2 c_z^2 s_x^2 \left\{c_z^2 + 2\right\}}{2\Delta_O^{'}} d_3k$$

Hence :

$$P_{x1} = -32 I_{13}^{(2)} \quad \text{(D6)}$$

**Relationship between $P_{x2}$ and $P_x^{(1)}$**

$$P_{x2} = \int \frac{16c_x^2 c_y^2 c_z^2 s_x^2 \left\{4c_y^2 c_z^2 + 3c_y^2 + 3c_z^2 - 4\right\}}{\Delta_O^{'}} d_3k = 8P_x^{(1)}$$

Hence :

$$P_{x2} = 8P_x^{(1)} = 64 I_{23}^{(2)} = -2P_{x1} \quad \text{(D7)}$$

**Relation ship between $P_{y1}^{(2)}$ and $P_x^{(1)}$**

$$2P_{y1}^{(2)} = \int \frac{4c_{2y} c_x^2 c_y^2 c_z^2 s_x^2 \left\{c_z^2 + 2\right\}}{\Delta_O^{'}} d_3k$$
$$= \int \frac{2c_{2y} c_x^2 c_y^2 c_z^2 s_x^2 \left\{c_z^2 + 2\right\}}{\Delta_O^{'}} d_3k + \int \frac{2c_{2z} c_x^2 c_y^2 c_z^2 s_x^2 \left\{c_y^2 + 2\right\}}{\Delta_O^{'}} d_3k$$
$$= \int \frac{2c_x^2 c_y^2 c_z^2 s_x^2 \left\{4c_y^2 c_z^2 + 3c_y^2 + 3c_z^2 - 4\right\}}{\Delta_O^{'}} = P_x^{(1)}$$

Hence :

$$2P_{y1}^{(2)} = P_x^{(1)} \quad \text{(D8)}$$

## Appendix E: Equality of diagonal elements $t_{11} = t_{22}$

| | |
|---|---|
| $t_{11} = -8.1344937604D-02$ | $t_{12} = -2.1305482540D-01$ |
| $t_{21} = -3.5493833258D-01$ | $t_{22} = t_{11}$ |

*Table E1. Average products tij's for the divacancy mechanism in the FCC lattice.*

### Reducing the expression of $t_{11}$

Starting from expression $-t_{11} = P^{(1)}_{\Sigma x} = P^{(1)}_{x} + P_{x1}pS^{(1)}_{y1} + W_O P_{x2} LL^{(1)}_{y2}$ while using Eqs. (16) and (D6-D7) yields:

$$-t_{11} = P^{(1)}_{x} - 32I^{(2)}_{13}\frac{I^{(1)}_{13}I_{22} - I^{(1)}_{23}I_{12}}{H} + 64I^{(2)}_{23}\frac{I^{(1)}_{23}I_{11} - I^{(1)}_{13}I_{21}}{H} \tag{E1}$$

where the denominator is defined by $H = I_{11}I_{22} - I_{21}I_{12}$.
Using Eqs. (B9) and (B12) transforms the numerator of the second term into $-I_{12}/8$ and the numerator of the third term into $(I_{11} - H)/8$. Hence the final compact expression:

$$-t_{11} = P^{(1)}_{x} + \frac{4I^{(2)}_{13}I_{12}}{\mathrm{H}} + \frac{8I^{(2)}_{23}(I_{11} - \mathrm{H})}{\mathrm{H}} \tag{E2}$$

### Reducing the expression of $t_{22}$

Starting from $-t_{22} = P^{(2)}_{\Sigma y1} + P^{(2)}_{\Sigma z1} = 2P^{(2)}_{\Sigma y1} = 2P^{(2)}_{y1} + 2P_{y11}pS^{(2)}_{y1} + 2W_O P_{y12} LL^{(2)}_{y2}$ and using Eqs. (16) and (D8) yields:

$$-t_{22} = P^{(1)}_{x} + 2P_{y11}\frac{I^{(2)}_{13}I_{22} - I^{(2)}_{23}I_{12}}{H} + 2P_{y12}\frac{I^{(2)}_{23}I_{11} - I^{(2)}_{13}I_{21}}{H}$$
$$= P^{(1)}_{x} + \frac{2I^{(2)}_{13}(P_{y11}I_{22} - P_{y12}I_{21})}{H} + \frac{2I^{(2)}_{23}(P_{y12}I_{11} - P_{y11}I_{12})}{H}$$

Using Eqs. (C6-C7) yields :

$$-t_{22} = P^{(1)}_{x} + \frac{2I^{(2)}_{13}(P_{y11}I_{22} - P_{y12}I_{21})}{H} + \frac{2I^{(2)}_{23}(P_{y12}I_{11} - P_{y11}I_{12})}{H}$$
$$= P^{(1)}_{x} - \frac{8I^{(2)}_{13}I_{21}}{H} + \frac{8I^{(2)}_{23}(1-H)}{H}$$

Using Eq. (B10) yields finally:

$$-t_{22} = P_x^{(1)} + \frac{4I_{13}^{(2)} I_{12}}{H} + \frac{8I_{23}^{(2)}(1-H)}{H} \quad \text{(E3)}$$